\documentclass[12pt]{article}
\usepackage{epsfig}

\begin{document}                                               
\begin{center}

\hspace{8cm}CLAS-NOTE 2009-016
\vspace*{1 cm}

{\bf Test Measurements of Prototype Counters for CLAS12 Central
Time-of-Flight System using $45$ MeV protons. } 

\vspace{4ex} 


\vspace*{1cm}
\underline{V.Kuznetsov$^{1,2}$, A.Ni$^{1}$, H.S.Dho$^{1}$, J.Jang$^{1}$, A.Kim$^{1}$, W.Kim$^{1}$} 
\vspace{2ex} 

$^{1}$ Kyungpook National University, 702-701, Daegu, Republic of Korea, \\
$^{2}$ Institute for Nuclear Research, 117132, Moscow, Russia.



\vspace{0.5ex} 

\end{center} 
\vspace{5mm}

\begin{abstract}
 
A comparative measurement of timing properties of magnetic-resistant fine mesh 
R7761-70 and ordinary fast R2083 photomultipliers is presented together with
preliminary results on the operation of R7761-70 PMs in magnetic field 
up to 1100 Gauss. The results were obtained using the proton beam 
of the MC50 Cyclotron of Korea Institute of Radilogical and Medical Sciences.

The ratio of the effective R7761-70 and R2083 TOF (or timing) resolutions was extracted by using
two different methods. The results are $1.05\pm 0.066$ and $1.07\pm 0.062$.
The gain of R7761-70 PMs is not affected by magnetic field. 
The R7761-70 TOF/timing resolution becomes $\sim 8\%$ better at 1100 Gauss 
if the external field is oriented parallel to the PM axis.
The results prove the advantages of the design of the CLAS12 Central
Time-of-flight system with fine-mesh photomultipliers in comparison 
with the ``conservative" design based on ordinary R2083 PMs and 
long bent light guides.   

\end{abstract}

\section{Introduction}

\begin{figure}
\vspace*{-0.5cm}
\epsfverbosetrue\epsfxsize=7.7cm\epsfysize=4.3cm\epsfbox{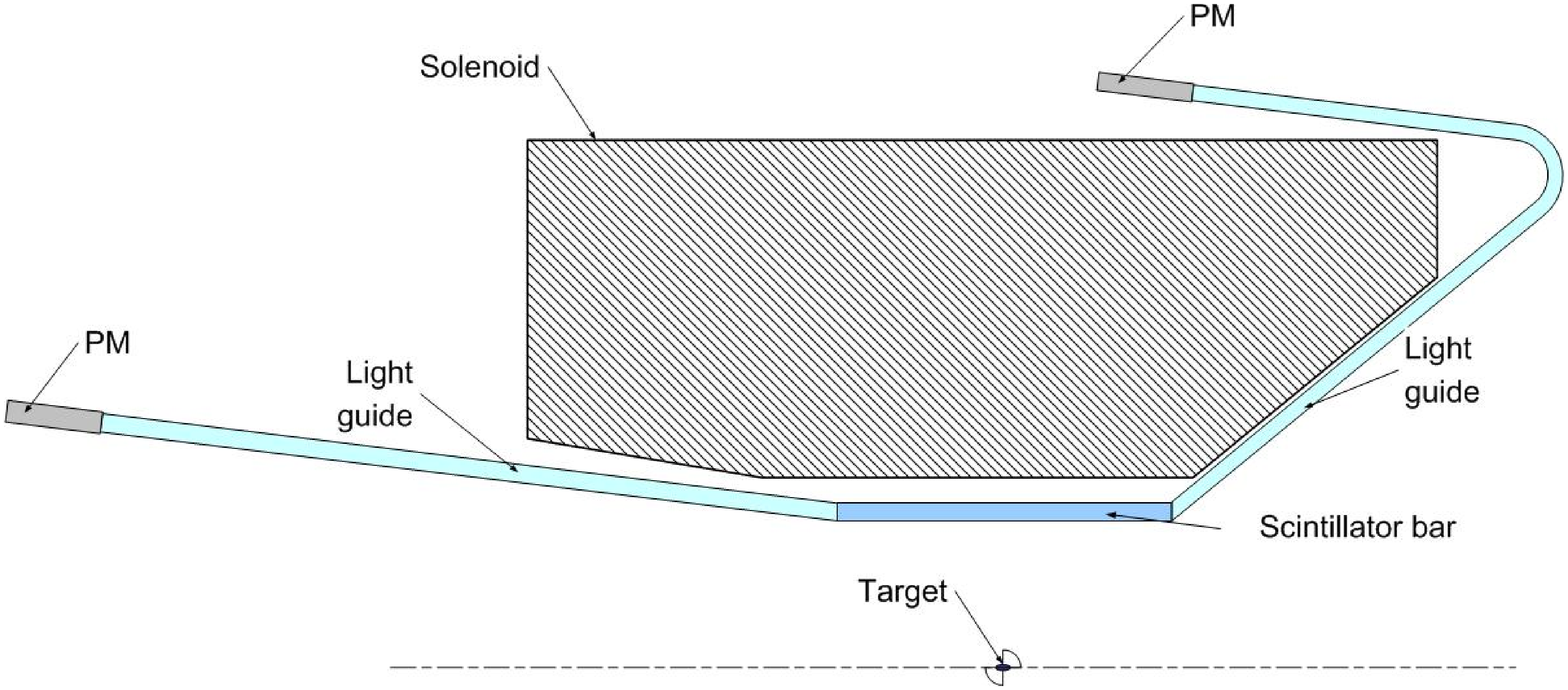}
\epsfverbosetrue\epsfxsize=5.7cm\epsfysize=3.9cm\epsfbox{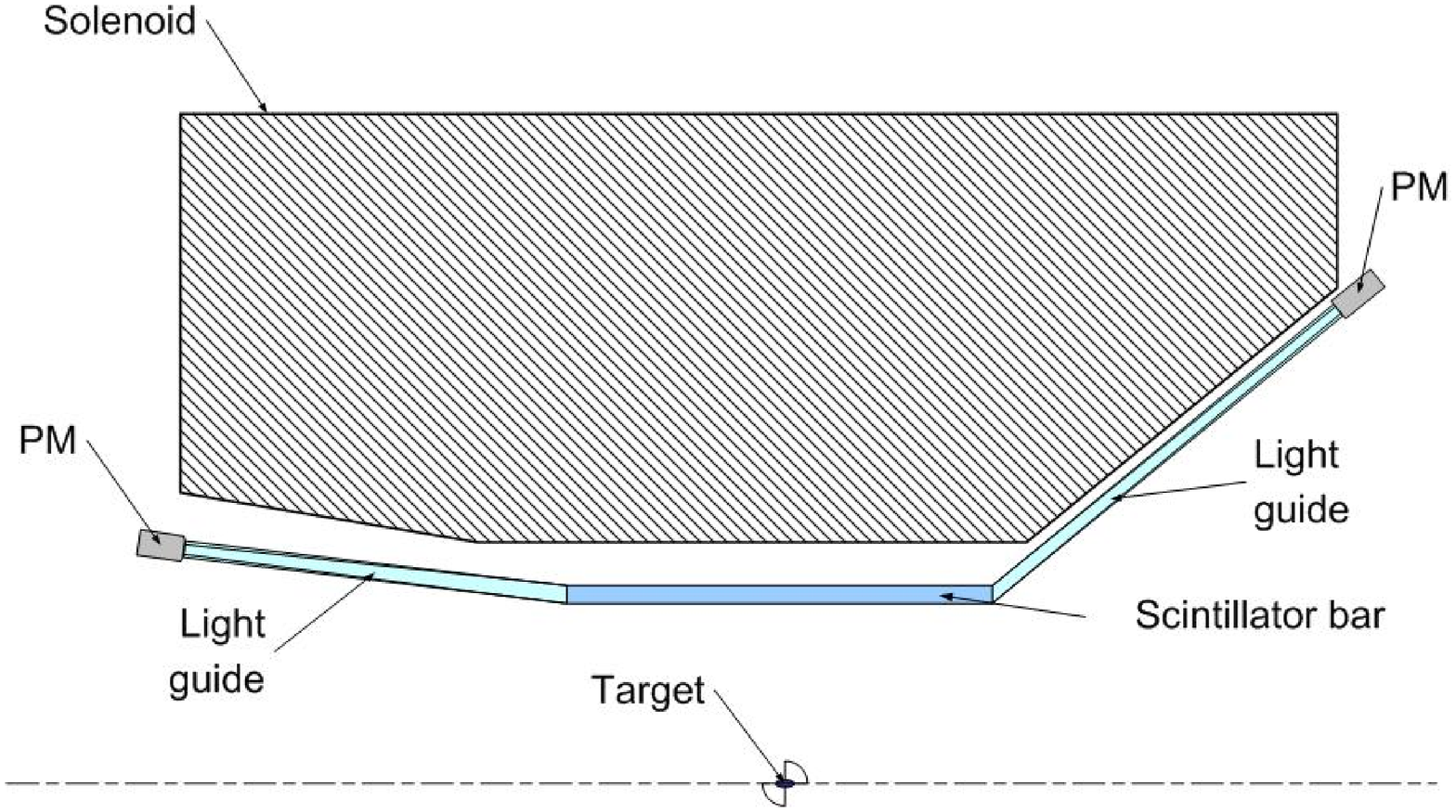}
\caption{Two versions of the CTOF design. On the left: ``conservative" design with long bent light guides and ordinary R2083 photomultipliers. On the right: CTOF design with fine-mesh photomultipliers
with short straight light guides.} 
\vspace*{-0.5cm} 
\label{fig:des1}
\vspace*{0.3cm}
\end{figure}

Initially the Central Time-of-Flight system (CTOF) was considered as 
a barrel made of $\sim50$ scintillator bars viewed by fast Hamamatsu R2083 
photomultipliers (PMs)  through $\sim1.5 - 1.7$ m long bent 
light-guides\cite{tdr,bat1} (left panel of Fig.~\ref{fig:des1}). 
The light guides are to deliver scintillation light to the regions outside 
the central solenoid where the magnetic field 
drops down to 300 - 1000 Gauss. Being properly shielded,
ordinary photomultipliers are expected to operate in this field.

Problems of this design are obvious:
\begin{itemize}
\item[-] The long and bent light guides would deliver a small portion 
of scintillation light to PMs thus deteriorating the time-of-flight (TOF) resolution;
\item[-] The mechanic construction would be rather complicate, ugly, and fragile. 
It would interfere with other CLAS12 sub-detectors;
\item[-] $\sim 100$ long light guides and $\sim 100$ magnetic shields would 
significantly increase the overall CTOF weight and costs;
\end{itemize}

The Nuclear Physics Group of Kyungpook National University, 
in collaboration with Inisitute for Nuclear Research, Moscow, suggests 
another solution based on magnetic-resistant fine-mesh photomultipliers. 
Fine-mesh photomultipliers can operate in magnetic field up to 1.5 Tesla. 
They could be placed closer to scintillator bars at positions 
where the magnetic field is 0.3 - 1 Tesla (right panel of Fig.~\ref{fig:des1}). 
The light guides would be much shorter, $\sim 0.50 - 0.8$ m long, and not bent.
No shields would be needed. The CTOF assembly would be simpler, 
more reliable, and less expensive.
A critical question is whether the acceptable TOF resolution could 
be achieved with fine-mesh of photomultipliers in comparison with R2083 PMs, 
and how it would be affected by the magnetic field 

There was some skepticism regarding fine-mesh photomultipliers. 
The authors of Ref.~\cite{bat1} wrote that ``...While the transition time 
spread (TTS) is compatible to TTS of R2083, its anode rise time (\textit{Rem.} of a 
fine-mesh photomultiplier) is 4 times larger. Therefore it is unlikely it may be used 
for precise timing...". Our consideration is different: 
the real PM anode rise time in a scintillation counter depends also 
on the light decay constant of a scintillator material. For Bicron-408 it is 2.5 ns - 
is much longer than the internal R2083 rise time.  
The most critical parameter is the transit time spread (TTS) which is in fact the
variation of electron propagation time inside a photomultiplier. 
The TTS of fine-mesh R7761-70 photomultipliers is $350$ ps. It is better than 
that of R2083 PMs ($370$ ps)~\cite{ham}.

The operation of fine-mesh photomultipliers in magnetic field 
was investigated in Ref.~\cite{it1,ru1,jp1}. The outcome of 
those studies was encouraging: the timing performance of fine-mesh 
photomultipliers is good and is not affected 
by magnetic field up to $\sim 0.8$ Tesla. 
The authors employed fast laser light pulses as a tool 
to study PM operation. This technique  is quite convenient 
but it doesn't allow to estimate the CTOF resolution and 
to judge between two CTOF designs.    

Until recently the cosmic-ray tracking was believed to be 
a main tool for the CTOF R\&Ds~\cite{bat1,bat}. The 
method uses three stacked parallel equidistant counters viewed by six phototubes under study. 
If a cosmic-ray muon crosses all three counters, the time (or coordinate) of a scintillation
in the middle counter is equal to a half of the sum of times/coordinates 
in the top and bottom counters\footnote{Detailed description of this 
method is available in \protect\cite{bat}}. This leads to the following relation 

\begin{equation}
\tau=\frac{1}{2}(t_{mid 1}+t_{mid_2})-\frac{1}{4}(t_{top 1}+t_{top 2}+t_{bot 1}+t_{bot_2})=0
\label{eq1}
\end{equation}
where $t_{top 1} .... t_{bot 2}$ denote 6 PM signal arriving times derived from 
TDC readouts. From Eq.~\ref{eq1} one may deduce the effective timing resolution 
in each PM channel 
\begin{equation}
\sigma_{PM}=\frac{2}{\sqrt{3}}\sigma(\tau)
\end{equation}
In practice the PM timing resolution is extracted from the width of 
the peak in the spectrum of $\tau$.

By employing this method, the KNU group measured the effective R7761-70 
timing resolution $\sigma_{PM R7761-70 \mu} \approx 52$ ps and found it similar
to that of R2083 $\sigma_{PM 2083 \mu} \approx 53$ ps ~\cite{knu1}.
The second result is in good agreement with the previous measurement~\cite{bat} 
$\sigma_{PM 2083 \mu}=52ps\pm 1$ ps. 

A disadvantage of this technique is that the measured PM arriving times 
$t_{top 1} .... t_{bot 2}$ are spoiled by time walks of 
constant-fraction discriminators (CFD).
Time walks depend on pulse shapes and especially on pulse heights. 
The CFD Phillips 715 and Ortec935 used in Ref.~\cite{bat1,bat,knu1}, generate  
time walks up to $\pm 50$ps~\cite{man}. 
Due to light attenuation the pulse heights of
muons events depend on hit coordinates and track angles.
The variation of the corresponding time walks is comparable with the measured 
PM timing resolution. That is why the results obtained with cosmic rays, 
require sophisticated corrections and are cut-,
electronic-, and analysis-dependent\footnote{The influence of time walks
on cosmic-ray results is discussed in detail in Ref.~\cite{knu1}.}.
They can be used only for rough preliminary estimates of the expected TOF resolution.

In this Note, we present another method to study the operation
of scintillation counters and photomultipliers
using a well-collimated proton beam. The method has been implemented at the 
MC50 Cyclotron of Korea Institute of Radiological and Medical Sciences.
Simultaneously we report the measurement of the TOF
resolution of a plastic-scintillation counter equipped with
fine-mesh Hamamatsu R7761-70 photomultipliers and compare it with that obtained with 
ordinary fast Hamamatsu R2083 PMs. We also report our first results on
the operation of fine-mesh R7761-70 photomultipliers in the magnetic field 
up to 1100 Gauss. 

\section{Fine-mesh photomultipliers}

Fine-mesh photomultipliers have been developed for high magnetic-field applications. 
Their dynode system has a structure of fine-mesh electrodes stacked in close proximity.
Such dynodes provide an improved pulse linearity and resistance 
to external magnetic field~\cite{ham}.

The survey of properties of fine-mesh photomultipliers is given in Table~\ref{tab0} together
with those of R2083 PMs. In general, the timing characteristics of 
the fine-mesh photomultipliers are worse: the anode rise time varies 
from $2.1$ to $2.7$ ns. For R2083 PMs this number is $0.7$ ns.
However the R7761-70 and R5505-70 transit time spreads 
are better than that of R2083 PMs.

\begin{table}
\begin{tabular}{lcccc}
Phototube& R2083 & R5505-70 & R7761-70 & R5924-70 \\
Dynode system & ordinary & fine-mesh & fine-mesh & fine-mesh \\
Photocathode dia (mm)& 39 & 17.5 & 27 & 39 \\
Photocathode type & bialkali & bialkali & bialkali & bialkali \\
Anode sensitivity (A/lm) & 200 & 40 & 800 & 700 \\
Anode rise time (ns) & 0.7 & 2.1 & 2.5 & 2.5 \\
Transit time (ns) & 16 & 5.6 & 7.5 & 9.5 \\
Transit time & 0.37 & 0.35 & 0.35 & 0.44 \\
spread (ns) & & & & 
\end{tabular}
\caption{Characteristics of fine-mesh photomultipliers in comparison with R2083.}
\label{tab0}
\end{table}

Due to geometrical dimensions only R7761-70 and R5924-70 PMs are suitable for CTOF.
Among them, R7761-70 PMs were chosen for initial tests because \\
i) Their timing characteristics are better than those of R5924-70; \\
ii) Small dimensions allow us to use them at the CTOF upstream ends; \\
iii) These phototubes are essentially cheaper than R5924-70 PMs. 

Hamamatsu Photonics offers H8409-70 assemblies. 
A H8409-70 assembly consists of a R7761-70 PM, a voltage divider
designed for positive HV supply, and a phototube housing. 
A positive divider necessarily includes a capacitor in the anode circuit
that separates the voltage supply from the anode output.
This capacitor may deteriorate timing properties and generate some 
level at high count rates. The geometric dimensions of H8409-70 assembly 
are larger than those of a single R7761-70 PM because of 
a phototube housing.
 
We developed our own voltage divider designed for negative HV. 
The distribution of potentials in the dynode system was optimized to achieve 
the best rise time. R7761-70 PMs were optically attached to scintillator bars 
and wrapped round with isolation tape without any housing. This made it possible 
to minimize the counter dimensions and to reduce bores of solenoids 
used in magnetic-field measurements (Section~\ref{sect:mf}). 


\begin{figure}
\vspace*{-0.5cm}
\epsfverbosetrue\epsfxsize=4.5cm\epsfysize=4.3cm\epsfbox{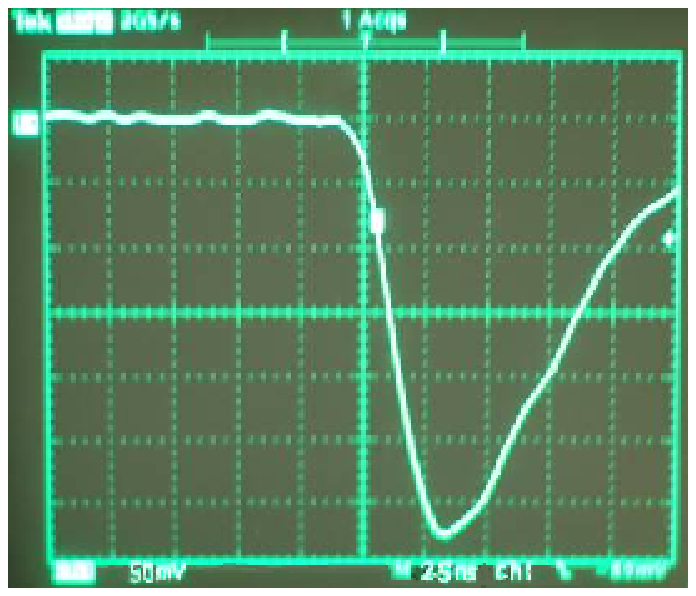}
\epsfverbosetrue\epsfxsize=4.5cm\epsfysize=4.3cm\epsfbox{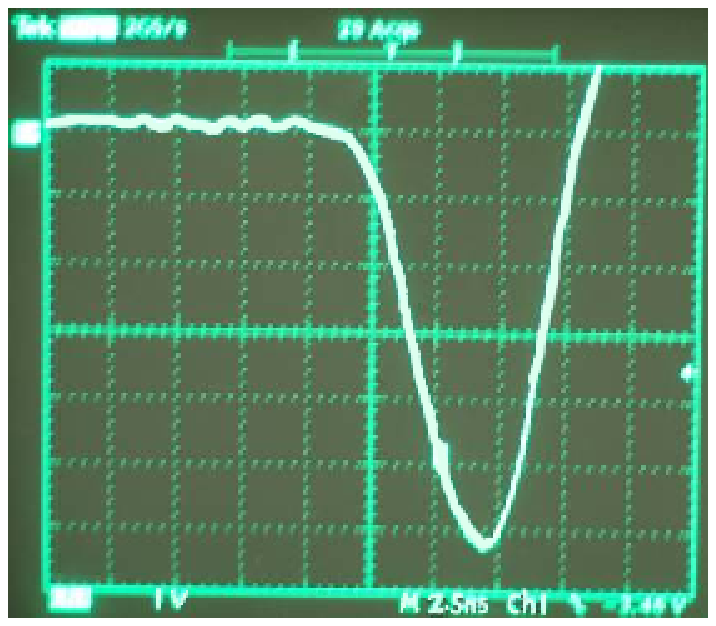}
\epsfverbosetrue\epsfxsize=4.5cm\epsfysize=4.3cm\epsfbox{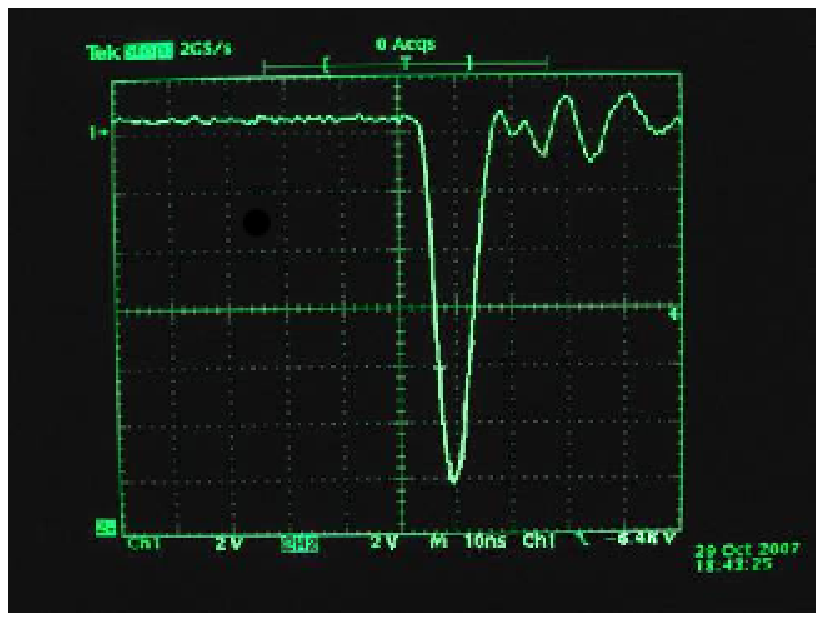}
\caption{Typical anode pulses from cosmic-ray muons obtained with R2083 (left) and 
R7761-70(middle and right).} 
\vspace*{-0.5cm} 
\label{fig:sig}
\end{figure}

Typical R7761-70 and R2083 signals are shown in Fig.~\ref{fig:sig}. 
They correspond to the detection of cosmic-ray muons in a 3 cm thick 
scintillator counter. The signals were obtained with $HV\sim 2200 V$ 
for R7761-70 PMs and $HV \sim 2500 $ V for R2083 PMs. The R7761-70 rise time $\sim3.2$ ns.  
It is only slightly worse than that obtained with R2083PMs ($\sim 2.5$ ns).
The R7761-70 pulse height reaches $12V/50 Ohm$. Such high pulse heights 
assure the operation of fine-mesh PMs in magnetic field 
in which the PM gain might be lower.

\section{Method and experimental setup}

If a long scintillator bar is viewed by two photomultipliers,
the PM arriving times $t_1$ and $t_2$ corresponding to a particle-induced 
scintillation are defined by the following relations:
\begin{equation}
t_1=TOF+x/v+Const\hspace*{1cm}t_2=TOF+(L-x)/v+Const
\end{equation}

\begin{figure}
\begin{center}
\includegraphics[width=10.0cm]{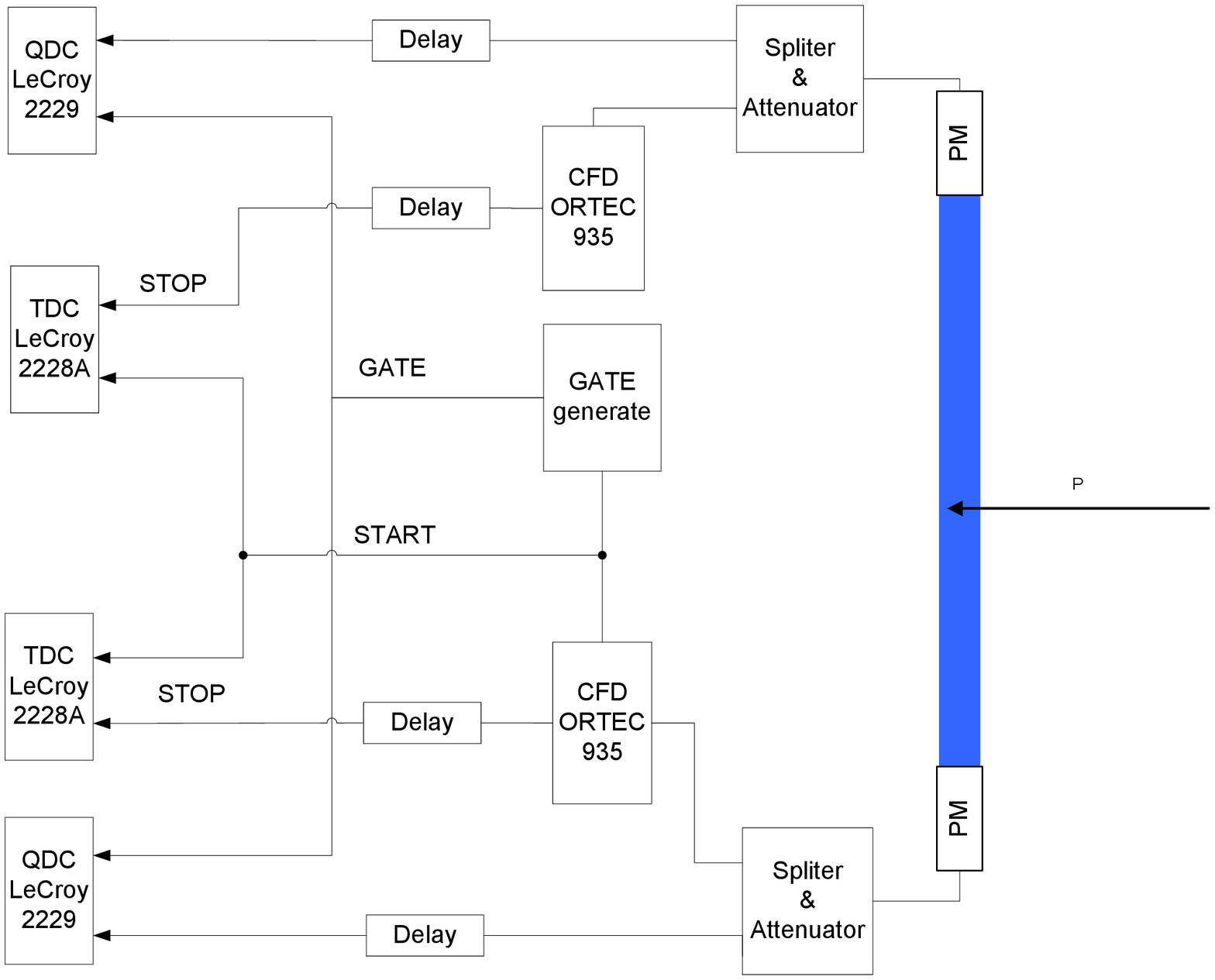}
\caption{Experimental setup.}\label{fig:scheme}
\end{center}
\end{figure}

\noindent where TOF is the time-of-flight of a particle from a certain point (target), 
$x$ is the coordinate of a scintillation  along the counter axis, $L$ is the total 
length of a bar, $v$ is the efficient speed of light propagation 
inside a bar, constants originate from cable and electronic delays.
Therefore TOF and the coordinate can be derived from the PM times
\begin{equation}
TOF=(t_1+t_2)/2+Const
\label{eq0}
\end{equation}
\begin{equation}
x=v(t_1-t_2)/2+Const
\label{eq5}
\end{equation}
The TOF resolution is
\begin{equation}
\sigma_{TOF}=\sigma((t_1+t_2)/2)=\frac{1}{2}\sqrt{\sigma^{2}(t_1)+\sigma^{2}(t_2)}
\end{equation}
where $\sigma(t_1)$ and $\sigma(t_2)$ are the effective 
timing resolutions in each PM channel. 
The variation of the time difference is equal
to the TOF resolution
\begin{equation}
\sigma((t_1-t_2)/2)=\frac{1}{2}\sqrt{\sigma^{2}(t_1)+\sigma^{2}(t_2)}=\sigma_{TOF}
\label{eq2}
\end{equation} 

If a scintillation counter is irradiated by a narrow proton beam, the beam generates 
a peak in the distribution of events over the coordinate $x$. 
As it follows from Eq.~\ref{eq5}, the coordinate distribution
is equivalent to the time-difference spectrum 
$(t_1-t_2)/2$ scaled by factor $v$. The width of this peak 
in the time-difference spectrum is defined by the size of the beam spot 
$\Delta x$ and by the timing performance of the counter
\begin{equation}
\sigma_{peak}=\sigma((t_1-t_2)/2+\Delta x/v)=\sqrt{\sigma_{TOF}^2+\sigma(\Delta x/v)^2},
\label{eq3}
\end{equation}
where $\sigma(\Delta x/v)$ denotes \textit{rms} in the coordinate distribution
of events due to the finite beam dimension. 
For a point-like beam $\Delta x \sim 0$ 
\begin{equation}
\sigma_{peak}\approx \sigma_{TOF}
\end{equation}
A well-collimated proton beam allows to derive the TOF resolution by measuring
the time-difference spectrum $(t_1-t_2)/2$.

In the practical implementation of this method a counter
made of the $50\times 3\times 2$ $cm^3$ Bicron-408 plastic scintillator bar
was used. The bar was viewed by  two photomultipliers under study. 

The counter was irradiated by the proton beam of the MC50 Cyclotron 
of Korea Institute of Radiological and Medical Sciences (KIRAMS). 
The beam was collimated by a collimator made of 3 stacked together 
7 mm thick steel plates. The plates had successively reducing holes
of  $5, 3,$ and $1$ $mm$ diameter in the first, second, 
and third plates respectively. 
The plate with the $1$ $mm$ dia hole was attached directly to the 2-cm side of the counter.
The whole collimation assembly and the counter were fixed at the end  of the beam pipe. 
As it was deduced from the Geant4 simulations, such collimator
reduces the contamination of protons scattered on the collimator walls and
minimizes the size of the beam spot.

The PM times $t_1$ and $t_2$ and their pulse heights 
$A_1$ and $A_2$ were digitized by LeCroy 2228B TDCs and LeCroy2229B QDCs 
and were recorded on-line (Fig.~\ref{fig:scheme}). One photomultiplier generated 
the common TDC START and triggered the acquisition.
Both photomultipliers generated the STOP signals for TDCs.

Typically, the beam current was set to $0.15$ nA.
At this current the count rate in the counter was $\sim 2\times 10^4 - 10^5$ Hz 
(Sect.~\ref{sect:cr}). The high voltages for fine-mesh R7761-70 PMs were set to
low values of $~1300 - 1350$ V, in order to fit QDC ranges.

\section{Beam of the MC50 Cyclotron}\label{sect:beam}
 
The MC50 Cyclotron of KIRAMS
has been built in 1985 for medical, nuclear-physics, and biological applications.
It can produce 20-50 MeV protons and deuterons. A neutron beam line is under 
construction. Beyond of a medical facility, there are three experimental hatches available 
for scientific research.
The beam spectrum and profile in each hatch depend on the accelerator adjustment,
the internal collimators, and other factors. They have to be adjusted for 
each experiment.

\begin{figure}
\centerline{\epsfverbosetrue\epsfxsize=14.0cm\epsfysize=17.0cm\epsfbox{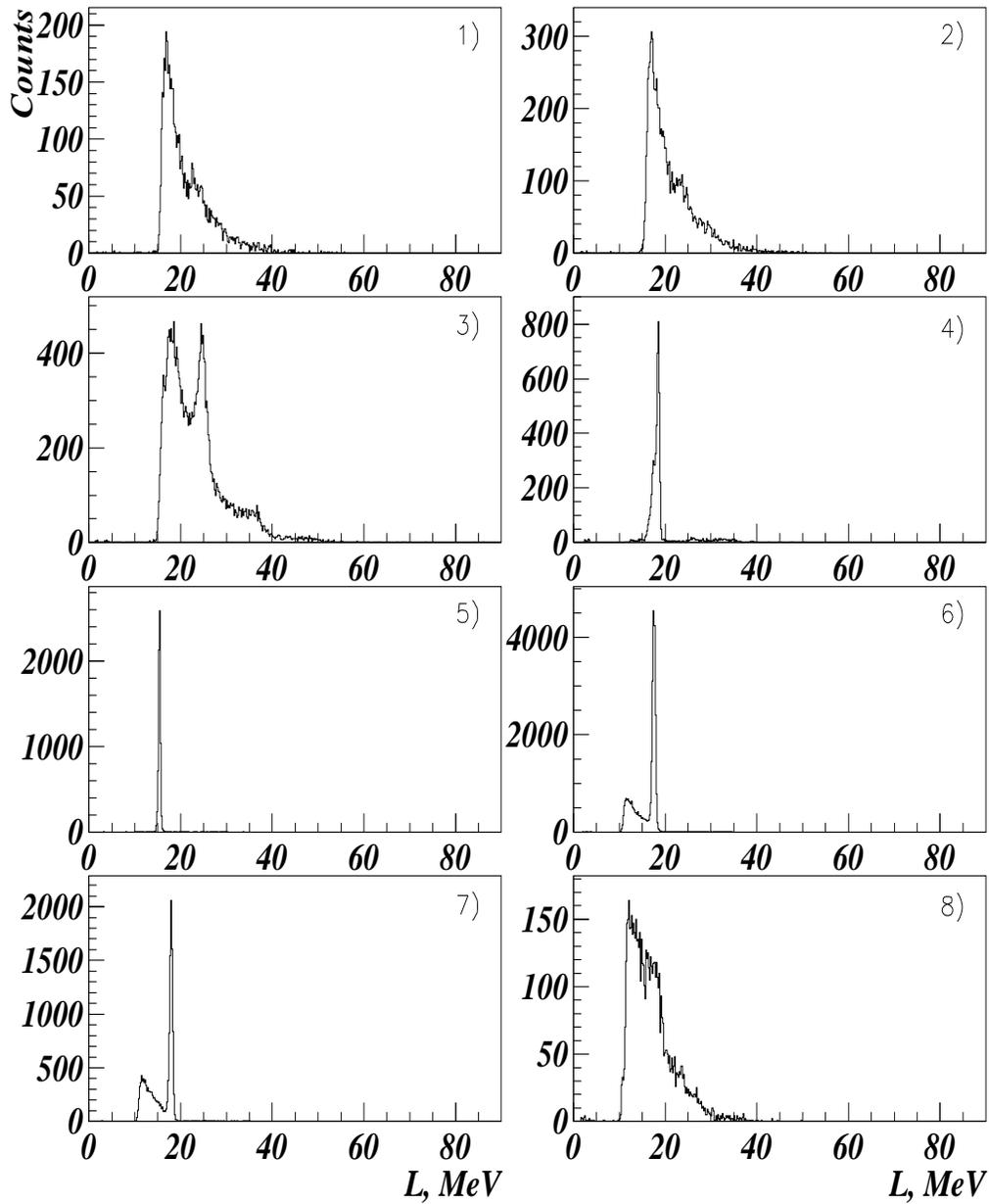}}
\caption{Procedure of beam adjustment. 8 beam spectra correspond to
8 successive steps in beam tuning. Calibration of the light output $L$ 
is explained in the Section~\ref{sect:calibr}. }\label{fig:badj}
\end{figure}

The data reported here were taken as a sequence of short ($\sim 1$ minute) 
measurements. In total there were 6 beam runs. Among them the data from three last 
runs are used in this Note. Each beam run lasted two-three days. 
It included apparatus installation at 
the MC50 beam line, calibrations, beam adjustment, and main measurements. 
The beam adjustment was a daily starting point and the most
complicated and time-consuming procedure.
The beam focusing and the position of the beam spot 
at the front end of the beam pipe were varied by the accelerator operator.  
The measured beam light-output spectrum was used as the criterion for 
the optimization of the beam quality. 
An example of this procedure is shown in Fig.~\ref{fig:badj}: 8 measured 
beam spectra correspond to 8 successive steps in the beam tuning. At the final point 
the set of beam parameters corresponding to the beam spectrum 
6 was chosen for the further running.

The adjusted beam spectrum (the panels 5 and 6 of Fig.~\ref{fig:badj}) exhibits
a main peak that corresponds to direct protons, and a tail at lower energies 
that is produced by particles which were either not properly accelerated or scattered 
inside the collimator.  
The position of the peak (i.e. the proton energy) depends on the operating parameters. 
Usually it was the same while in some cases the  peak position
was rather different. Such spectrum is shown in the 
panel 6 of Fig.~\ref{fig:badj}. There the peak position is nearly 
twice lower than its normal value.
Other beam spectra from the runs used in this report
are shown in Fig.~\ref{fig:bsp1},~\ref{fig:bsp2}. 
The calibration of these spectra is explained in the section~\ref{sect:calibr}.

\section{Birks' effect and light-output calibration}\label{sect:calibr}.

Plastic scintillators do not respond linearly to the ionization density.
Very dense ionization columns emit less light  than that expected on the basis of 
$\frac{dE}{dx}$ for minimum ionizing particles. 
This non-linear response of scintillators is called Birks' effect~\cite{birks}.   
Due to Birks's effect, protons which stop inside 
a detector, produce less light per unit of deposited energy than 
relativistic minimum-ionizing particles.
 
The semiempirical Birks' law is 
\begin{equation}
\frac{dL}{dx}=L_{0}\frac{\frac{dE}{dx}}{1+k_b\frac{dE}{dx}}
\label{eq:birks}
\end{equation}
\noindent where $L$ is the scintillator light production, $L_0$ is the specific light 
production at low ionization densities (i.e. the light produced by a relativistic
minimum-ionizing particle per a unit of deposited energy),
 $x$ is a coordinate along the particle track inside a scintillator volume,
and $k_b$ is Birks' constant which must be determined for each scintillator
experimentally. An interpretation of Birks' effect
was proposed by C.Chou~\cite{chou}. He corrected Birks' formula as
\begin{equation}
\frac{dL}{dx}=L_{0}\frac{\frac{dE}{dx}}{1+k_b\frac{dE}{dx}+k_c(\frac{dE}{dx})^2}
\end{equation}
\noindent where $k_b$ and $k_c$ are adjustable constant.

One can write Birks' law in the general form
\begin{equation}
\frac{dL}{dx}=k_p(E)L_{0}\frac{dE}{dx}
\label{eq:bic1}
\end{equation}
\noindent where $k_p(E)$ is the Birks's coefficient that depends on $E$, on
scintillator material, and on the type of a particle.



Low-energy particles generate less light than 
relativistic minimum-ionizing (MI) particles. With the increase of 
the particle energy ($\beta \to 1$),
the ratio of light production to deposited energy 
asymptotically approaches a constant value $L_0$
which is the same for all particles. The latter is clear, 
for example, from the bi-dimensional plots light-output vs TOF 
for charged pions and protons obtained
with the forward lead-scintillator TOF wall~\cite{rw} at the GRAAL facility.
It is convinient to assign this value to $L_0=1$. 
In this case the energy deposited by a minimum-ionizing 
particle in a detector volume can be used as a measure for light output.

TOF resolution depends on the number of photoelectrons produced 
at PM photocathodes. The latter is proportional to light output.
The CTOF R\&D requires to extrapolate the results obtained with 
the MC50 protons to those expected for fast minimum-ionizing particles.
That is why the proton light output was calibrated 
by assigning it to the light output produced by high-energy
cosmic-ray muons. 

\begin{figure}[t!]
\begin{center}
\includegraphics[width=10.0cm]{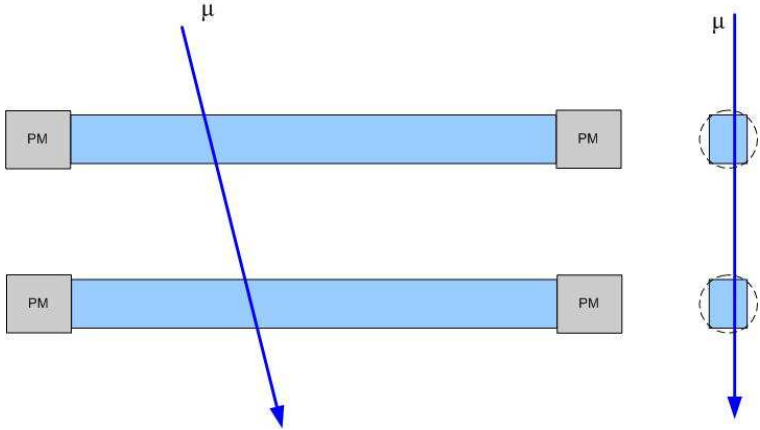}
\caption{Schematic two-side view of the calibration measurement with cosmic-ray muons.}
\label{fig:mmeas}
\end{center}
\end{figure}

The muon spectrum was measured just before and/or just after beam measurements. 
Schematic view of this measurement is shown in Fig.~\ref{fig:mmeas}.
The counter was disattached from the beam pipe and 
turned at $90^{\circ}$ such that the effective counter thickness of 3 cm 
was the same for the muons and the beam protons.  
The count rate of muon events is essentially lower than the 
background in the experimental hutch.  
To reject this background, the second similar counter
was placed $35$ cm below the main counter.
The coincidence of four signals from two counters was requested 
to trigger the acquisition. This made it possible to select on-line mostly
those events in which a cosmic-ray muon passes both counters (Fig.~\ref{fig:mmeas}). 

Beam protons stop inside the counter and totally deposit their energies
in the counter volume. Fast muons pass through
the counter and act as minimum-ionizing particles. Their energy 
depositions depend on the track length inside the counter, i.e. on 
the angle between the muon trajectory 
and the counter surface.

\begin{figure}
\vspace*{-0.5cm}
\epsfverbosetrue\epsfxsize=6.7cm\epsfysize=6.7cm\epsfbox{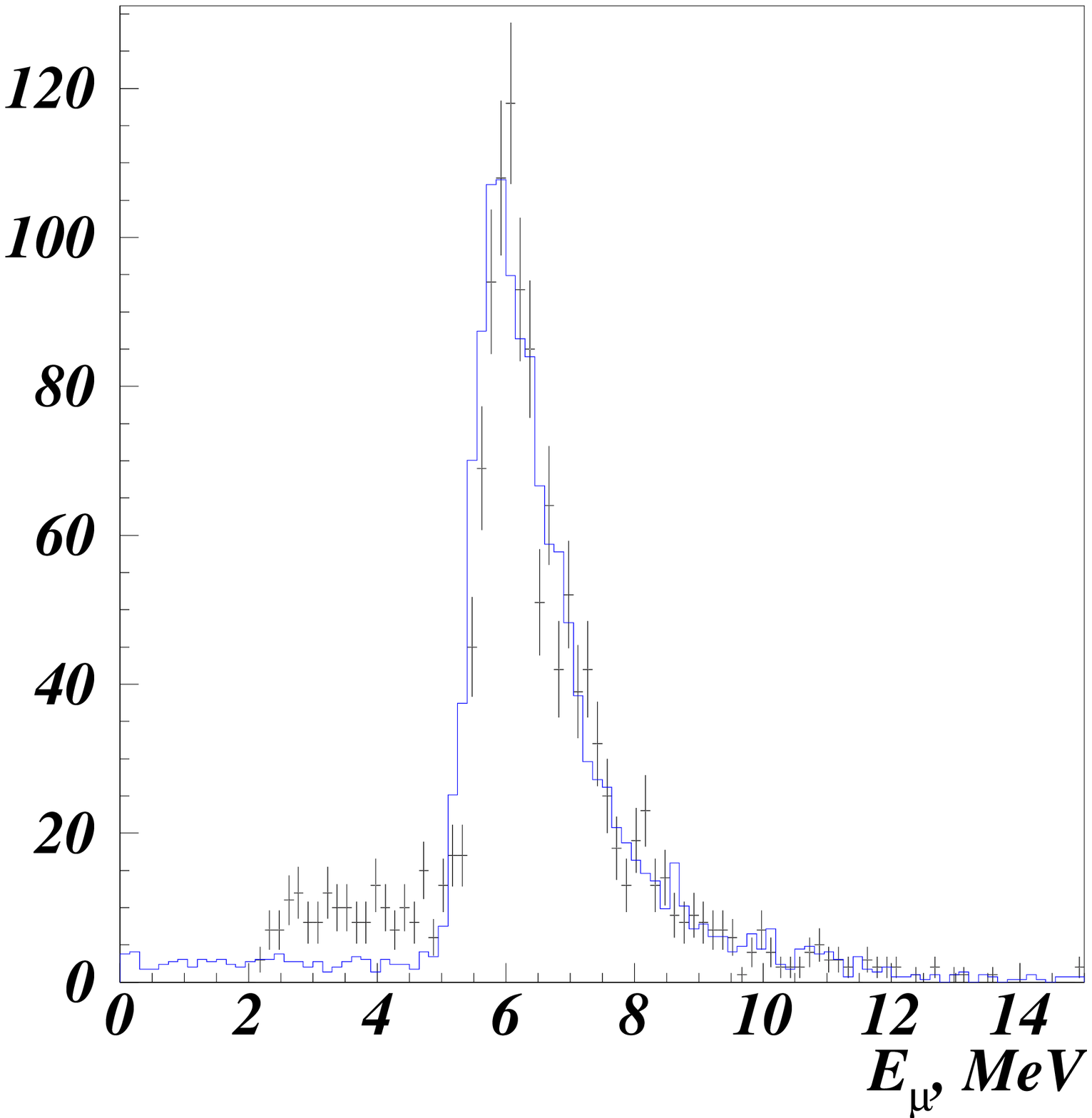}
\epsfverbosetrue\epsfxsize=6.7cm\epsfysize=6.7cm\epsfbox{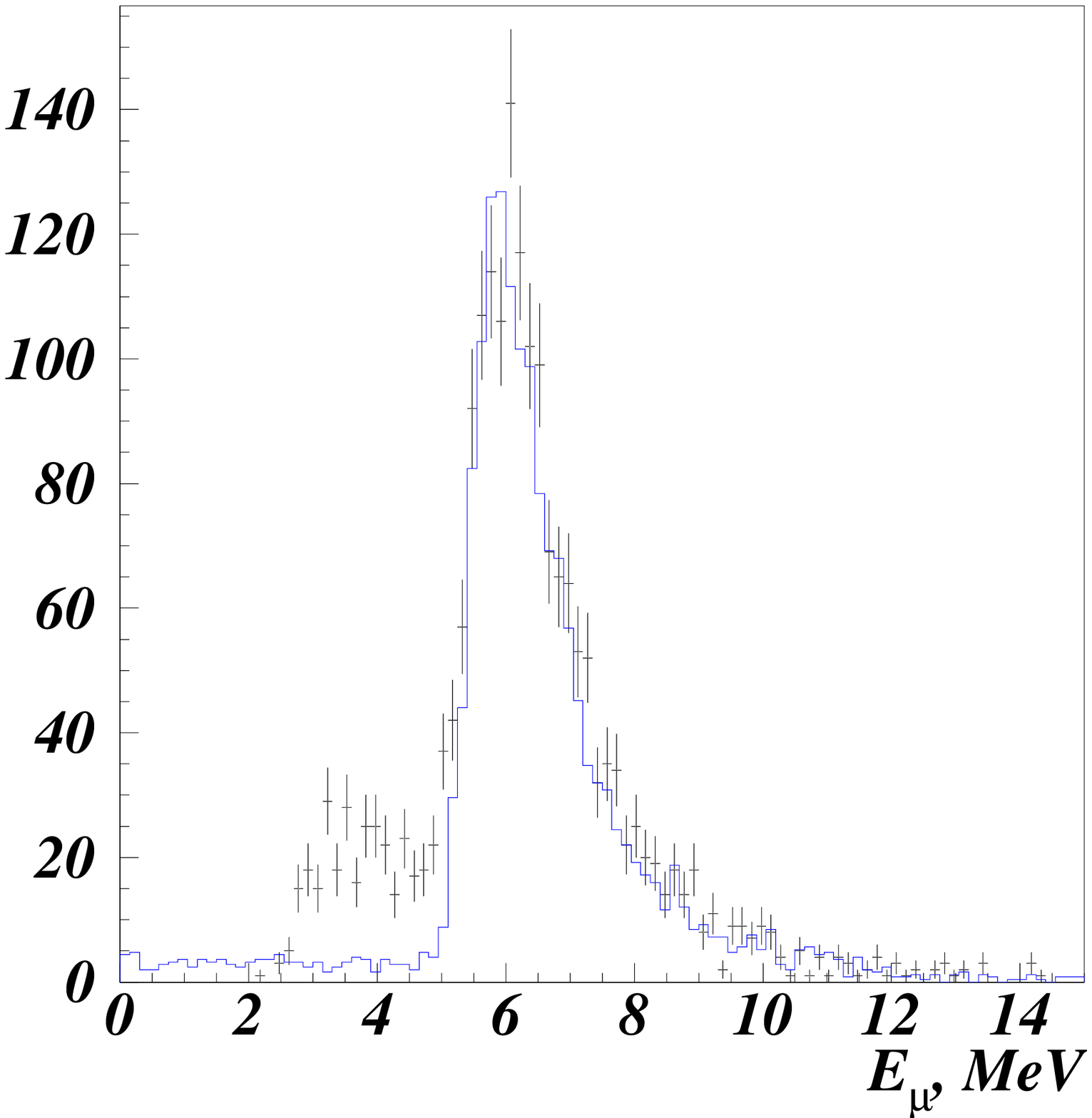}
\caption{Calibrated 
muon light-output spectra (Data from January 6, 2009). 
On the left: the spectrum obtained with 
fine-mesh R7761-70 photomultipliers. On the right: the spectrum obtained 
with ordinary R2083 photomultipliers. Solid lines are simulated spectra
of energy deposited by muons. 
} 
\vspace*{-0.5cm} 
\label{fig:calibr}
\end{figure}

Two samples of the muon spectra are shown in Fig.~\ref{fig:calibr}. They contain 
the asymmetric peak with the maximum at $\sim 6$ MeV. This maximum corresponds
to the energy deposited by a minimum-ionizing particle
crossing the counter in the perpedicular direction. The higher-energy part
is formed by those muons which cross the counter
at smaller angles.

The light output can be derived from the QDCs readouts $A_{1}$ and $A_{2}$ 
after the pedestal subtraction as 
\begin{equation}
L = C\sqrt{(A_1-ped_1)(A_2-ped_2)}
\end{equation}
\noindent where $C$ is the calibration coefficients that relates
QDC channels to the light output.
The pedestals were measured by delaying the QDC gates for $~200$ ns.
The calibration coefficient $C$ was obtained  by comparing
the measured muon spectrum with the simulated 
spectrum of the energy deposited by muons (Fig.~\ref{fig:calibr}).

As a cross-check, this calibration was applied to the data collected in the same 
beam run (January 6, 2009) with two counters. One counter was equipped 
with fine-mesh photomultipliers and the other with ordinary R2083 PMs.
The data were collected one-by-one keeping fixed the beam tuning.
The muon spectrum with R7761-70 photomultipliers was collected just before 
the beam run while the one with R2083 PMs was taken immediately after. 
Both muon spectra are shown in Fig.~\ref{fig:calibr}.
The calibration coefficients extracted from the muon spectra were then used 
to reconstruct the beam spectra. 

\begin{figure}
\vspace*{-0.5cm}
\epsfverbosetrue\epsfxsize=6.7cm\epsfysize=6.7cm\epsfbox{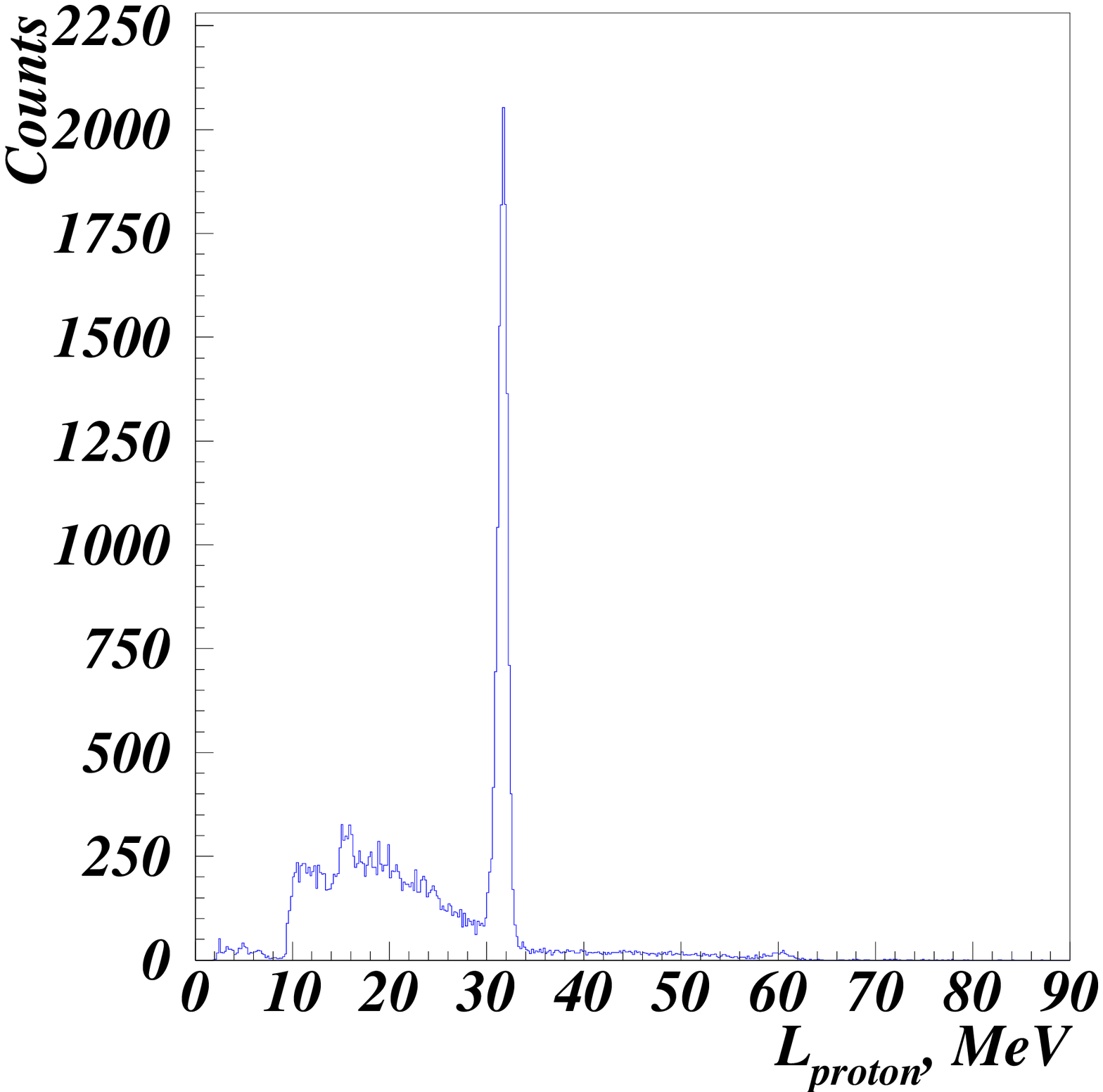}
\epsfverbosetrue\epsfxsize=6.7cm\epsfysize=6.7cm\epsfbox{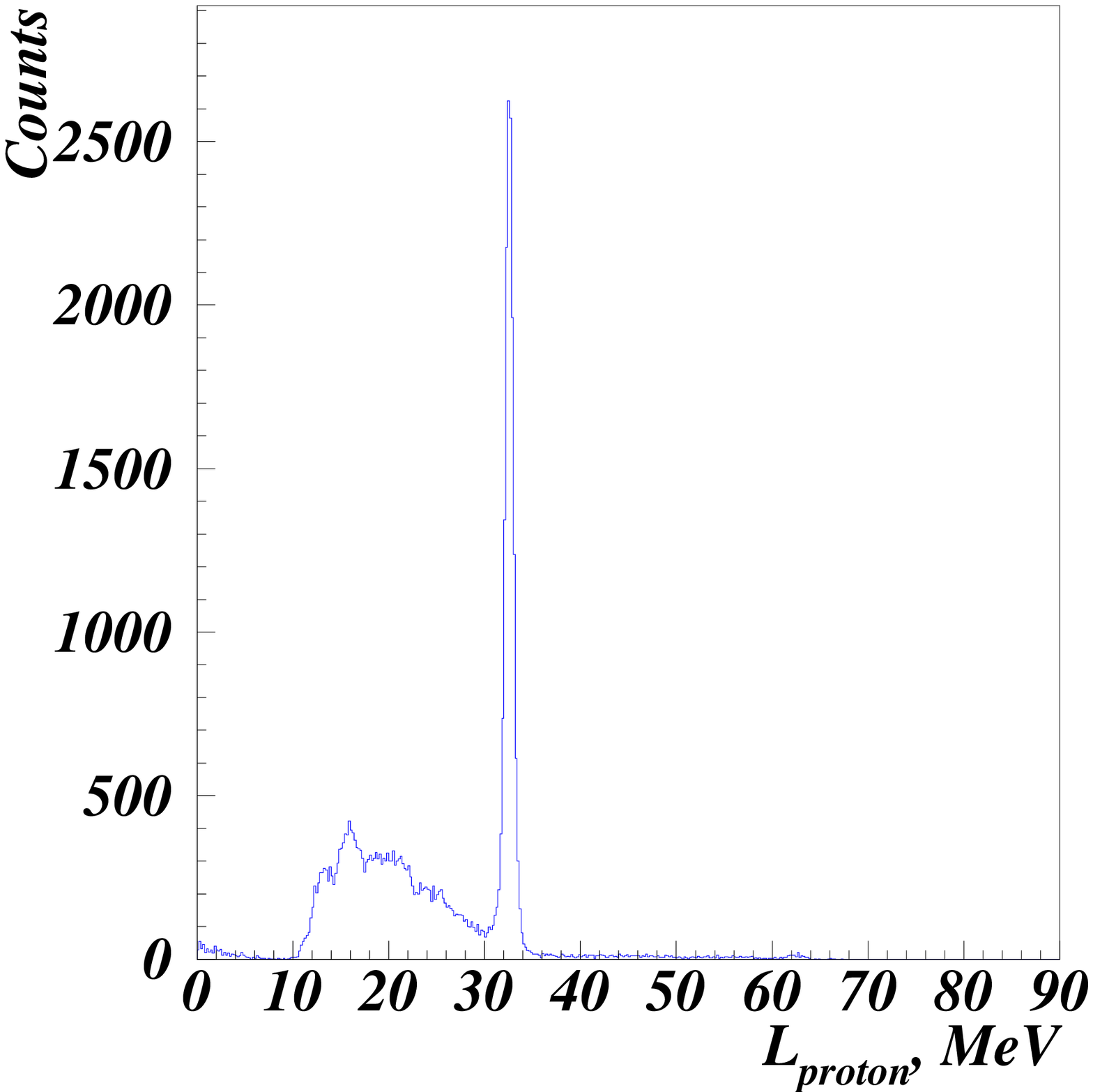}
\caption{Beam spectra measured with the two different counters
(Data from January 6 2009). 
On the left: the spectrum obtained with 
fine-mesh R7761-70 photomultipliers. On the right: the spectrum obtained 
with ordinary R2083 photomultipliers.} 
\vspace*{-0.5cm} 
\label{fig:bsp1}
\end{figure}
   
The results are shown in Fig.~\ref{fig:bsp1}. 
The beam peaks are located at the similar positions near 31.5 MeV. 
This proves the quality of the calibration procedure.
The value of 31.5 MeV corresponds to the light output produced
by a  minimum-ionizing particle that deposit $\Delta E = 31.5$ MeV in 
the counter volume. The beam protons stop inside the counter and deposit 
more energy, but, due to Birks' effect,
generate the same light output.  
 
The position of the peak in the spectrum taken next day 
(January 7 2009) is essentially lower, about $\sim 17.5 $ MeV 
(the panel 6 of Fig.~\ref{fig:badj}). This fact illustrates 
the importance and the influence of the MC50 beam tuning.

\begin{figure}
\vspace*{-0.5cm}
\epsfverbosetrue\epsfxsize=6.7cm\epsfysize=6.7cm\epsfbox{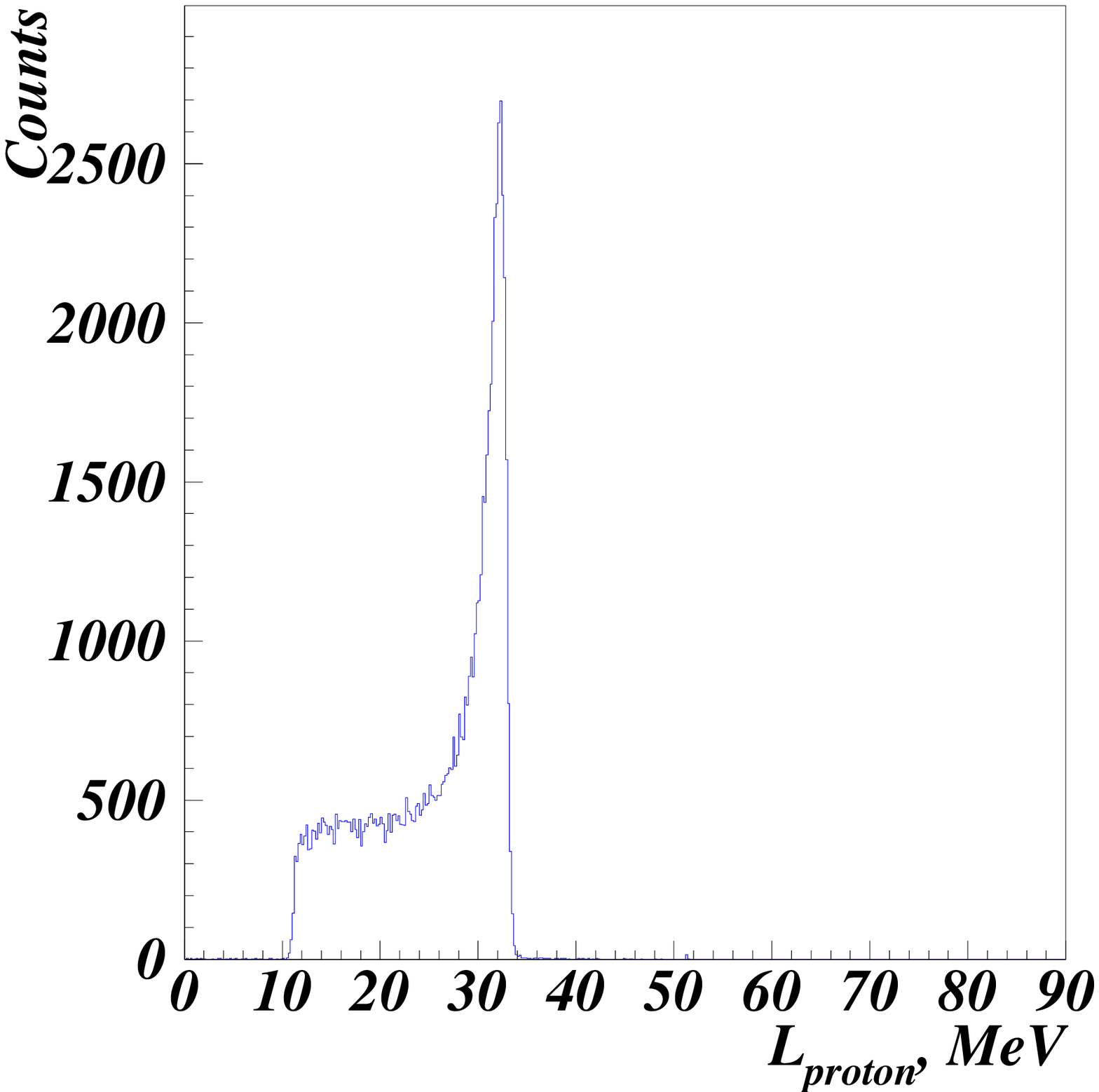}
\epsfverbosetrue\epsfxsize=6.7cm\epsfysize=6.7cm\epsfbox{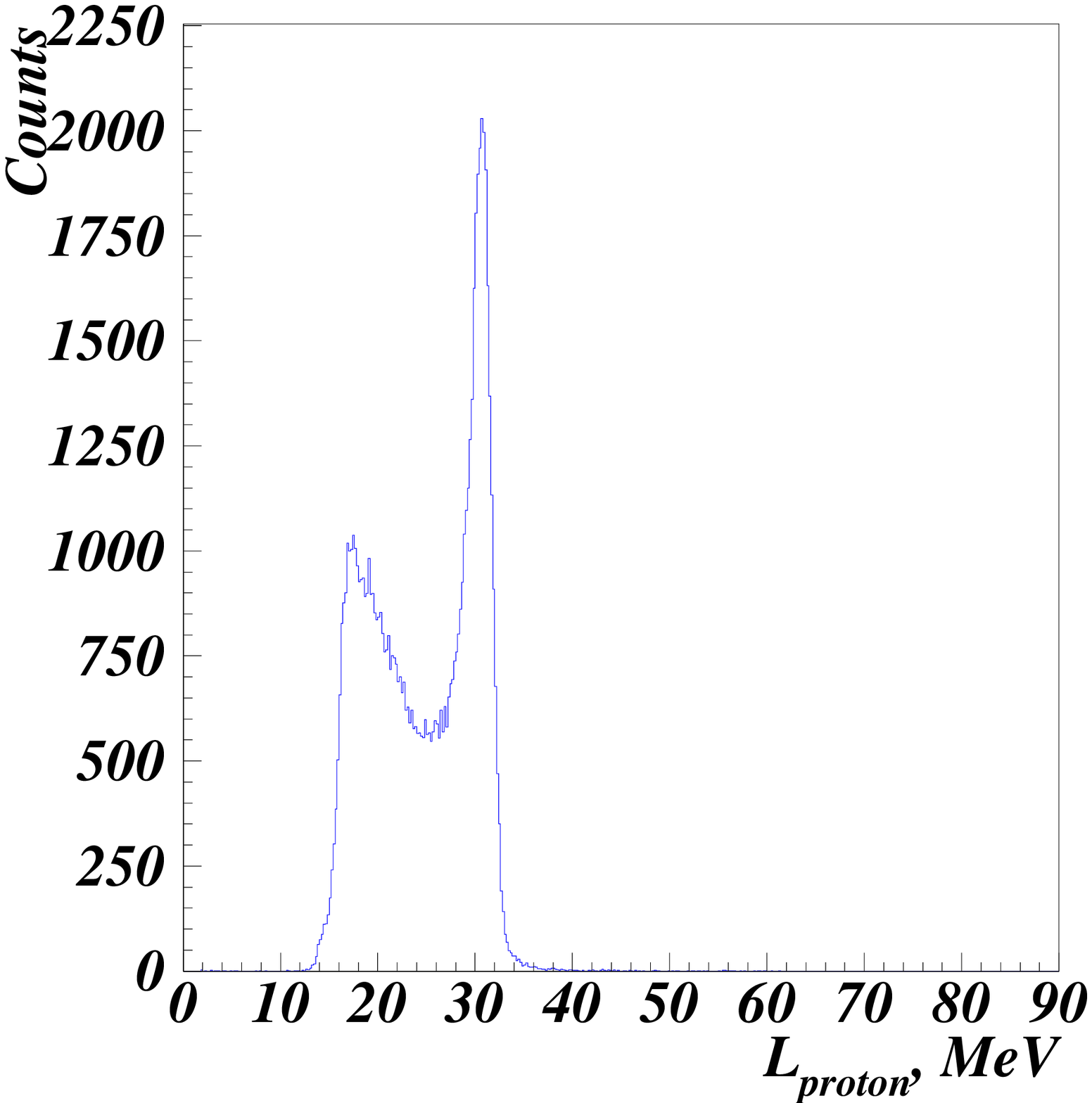}
\caption{Beam spectra measured in October and November 2008. 
On the left: the spectrum obtained with 
fine-mesh R7761-70 photomultipliers (October 2008). 
On the right: the spectrum obtained with ordinary R2083 photomultipliers (November 2008).} 
\vspace*{-0.5cm} 
\label{fig:bsp2}
\end{figure}

The other beam spectra from the beam runs used in this report (October and November 2008)
are shown in Fig.~\ref{fig:bsp2}.
The peaks are located at $L\approx 31.5$ MeV. However the shapes of these spectra
are different: the peak is wider and is smoothly continued by the low-energy tail.
The reason for this difference still has to be understand. 
Our assumption is the internal MC50 beam collimator was
removed during October/November 2008 beam runs. The data from October/November 2008 
were found suitable to retrieve the dependence of the time-of-flight 
resolution on the light output (Sect.~\ref{sect:rlout}).

\section{Count rate}\label{sect:cr}

\begin{figure}
\vspace*{-0.5cm}
\epsfverbosetrue\epsfxsize=6.7cm\epsfysize=6.7cm\epsfbox{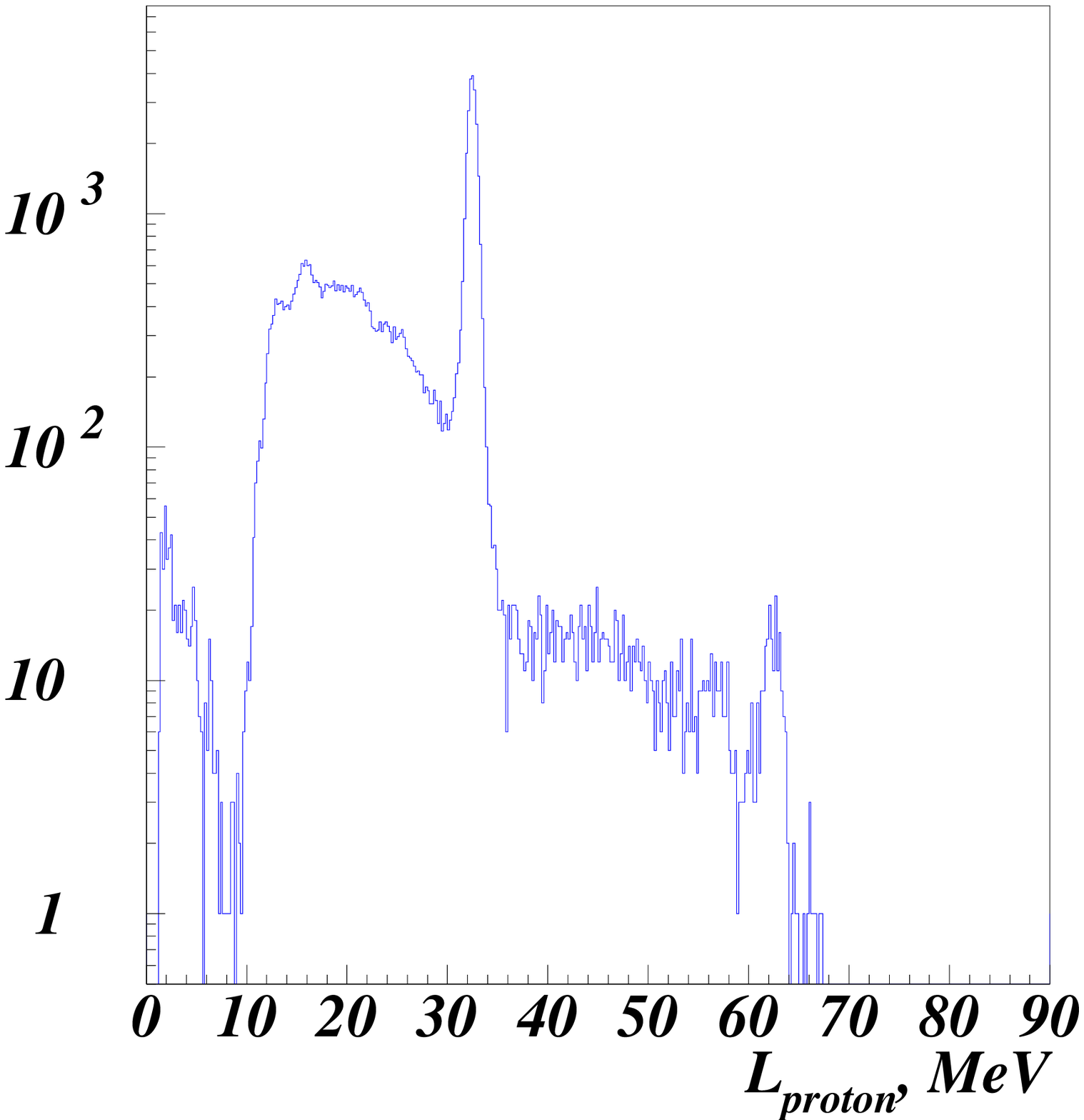}
\epsfverbosetrue\epsfxsize=6.7cm\epsfysize=6.7cm\epsfbox{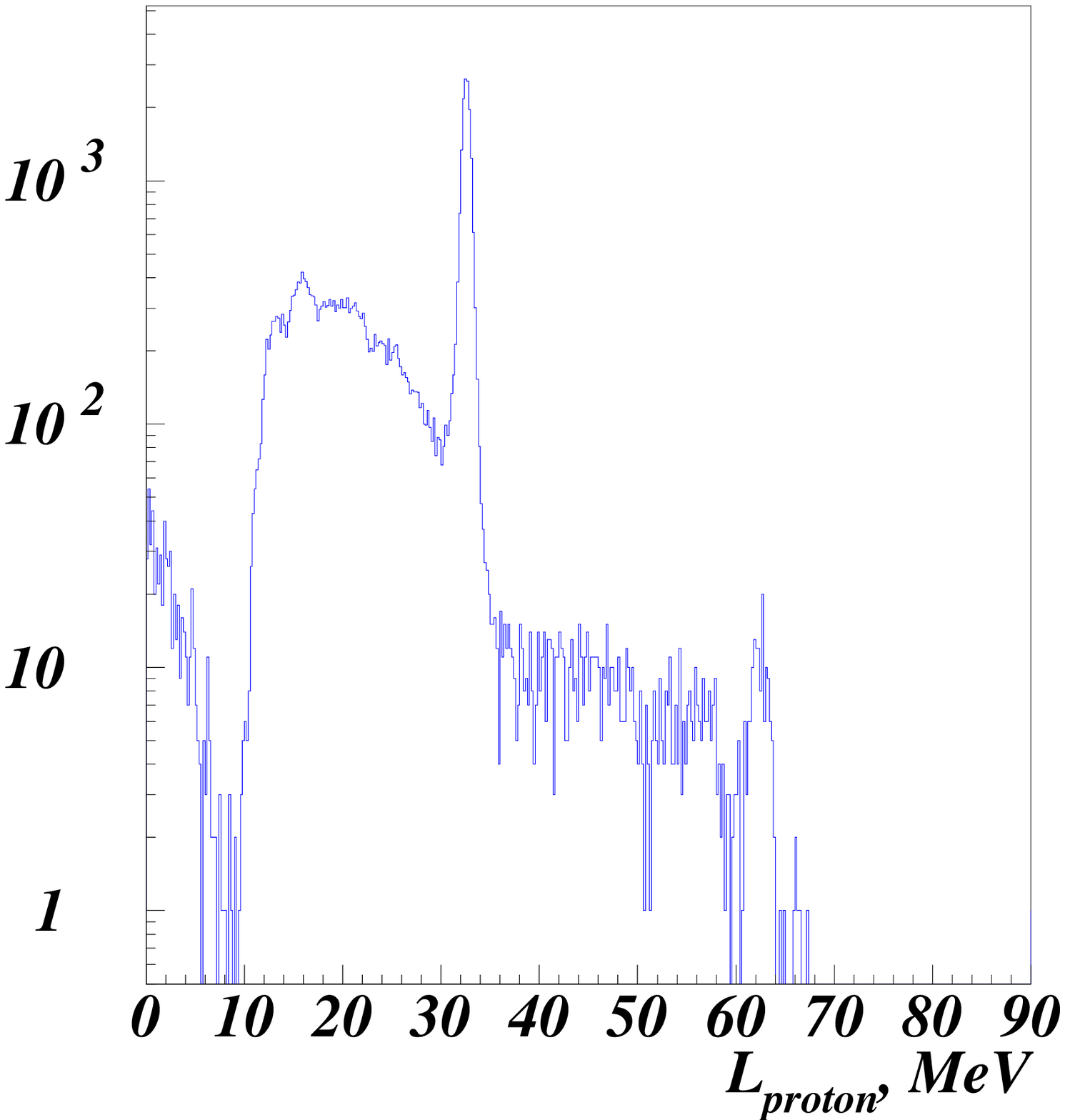}
\caption{Beam spectra in logarithmic scale.
The legend is the same as in Fig.~\protect\ref{fig:bsp1}.} 
\vspace*{-0.5cm} 
\label{fig:bsp3}
\end{figure}

The count rate in the counter depends on the beam current and tuning.
In the measurements presented in this Note, the beam current was set to its lowest possible 
value $0.15$ nA. The count rate can be estimated from the contamination of pileups in 
the beam spectrum. The probability that two pulses are encoded by QDC as one is 
\begin{equation}
P_{pileup}=\frac{N_{pileup}}{N_{tot}}=wN_{cr}
\label{eq:pp}
\end{equation}
\noindent where $N_{pileup}$ is the number of pileup events in a beam spectrum,
$N_{tot}$ is the total number of events, $w$ is the width of the QDC gates 
($100$ ns in our case), and
$N_{cr}$ is the count rate in the counter.

Fig.~\ref{fig:bsp3} shows a typical beam spectrum in the logarithmic scale. 
It contains events whose pulse heights are higher than the peak position. 
These are pileups. The fraction of such events typically varies from
0.2 to 1\%.  Following Eq.~\ref{eq:pp}, one may estimate the count rate as
\begin{equation}
N_{cr}\approx 2\times 10^4 - 10^5 Hz
\end{equation}

At this count rate and the high voltage $\sim 1300$V the average anode 
current of R7761-70 photomultipliers was in the range of $1.6 - 8 \mu A$. 
Some tentative measurements were carried out at higher count rate/average anode 
current up to  $5\times10^5$ Hz/$\sim 40 \mu A$.

It is worth noting that Ref.~\cite{it1} quotes the maximum average anode
current (typically $100 \mu A$ for R5024-70 PMs ) which must not be exceeded. 
Beyond this limit the PM gain is expected to sharply drop down.
For those two R7761-70 samples used in our tests, no 
reduction of the gain at the anode current up to $40 \mu A $
was observed. More comprehensive tests at high count rates are planned for 2009. 

\section{Getting TOF resolution}\label{sect:tof}

\begin{figure}
\begin{center}
\includegraphics[width=13.0cm]{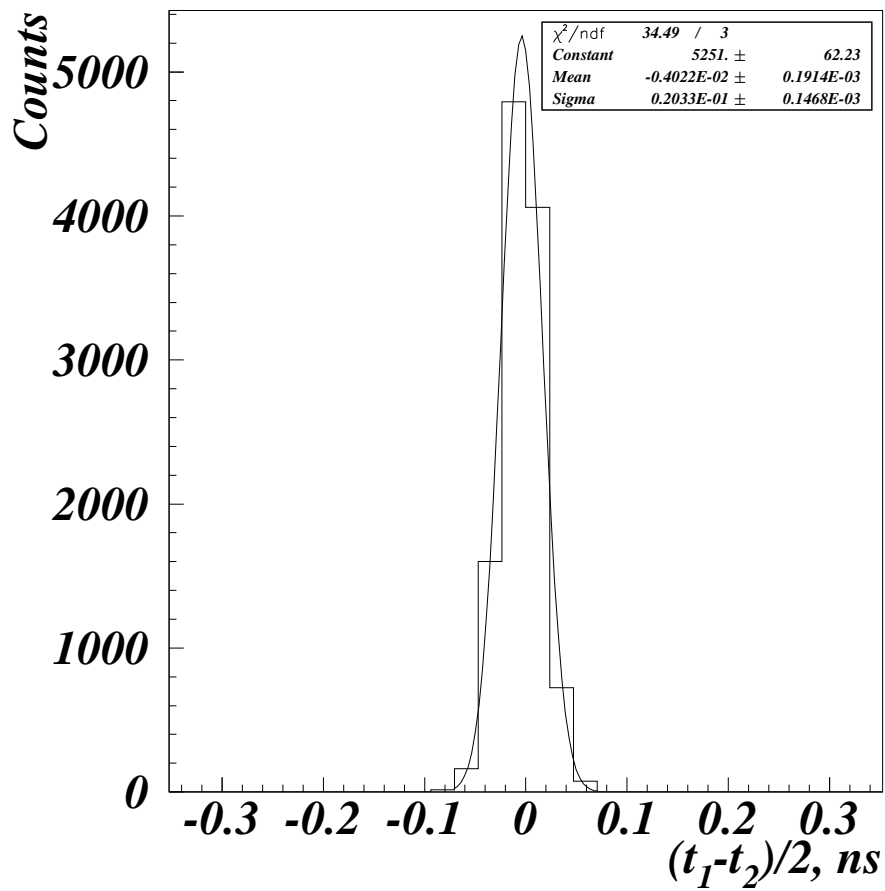}
\caption{Example of $(t_1-t_2)/2$ spectrum.  
Gaussian fit of the peak is one of the ways
to extract the TOF resolution.}
\label{fig:tres1}
\end{center}
\end{figure}

An example of the measured $(t_1-t_2)/2$ spectrum is shown in Fig.~\ref{fig:tres1}. 
The spectrum contains a narrow peak generated by the beam protons.
The width of the peak can be extracted by means
a Gaussian fit of this peak. However, because of the limited TDC resolution of 
$\sim 47$ ps/ch, the result is quite sensitive to 
the histogram binning. We found critical to set the width of
the histogram bins equal to the discreetness of $(t_1-t_2)/2$, i.e.
equal to a half of the width of one TDC channel $47 ps/2$. 

To cross-check the results, another method was employed as well.  
The center of the peak gravity $M_{peak}$ 
and the mean square deviation $\sigma_{peak}$ were directly calculated
from the sample of collected events as
\begin{eqnarray}
M_{peak}=\frac{1}{N_{ev}}\sum_{i=1}^{N_{ev}}x_{i} \\
\sigma_{peak}=\sqrt{\frac{1}{N_{ev}}\sum_{i=1}^{N_{ev}}(M_{peak}-x_i)^2},
\end{eqnarray}
\noindent where $N_{ev}$ denotes the total number of selected events,
$x_{i}$ is the time difference derived from the TDC readouts $ch_{tdc1_i}$ and $ch_{tdc2_i}$
for each recorded event as
\begin{equation}
x_{i}=r(ch_{tdc1_i}-ch_{tdc2_i}) .
\end{equation}
\noindent Here $r$ denotes the TDC scale (i.e. the width of one TDC channel). 
In the data analysis it was fixed to $r=47 ps/ch$. In reality $r$ is 
affected by the TDC differential non-linearity.
For the LeCroy2228B TDC unit used in these measurements, 
it was measured and found varying from $45$ to $49 ps/ch$. 
This variation leads to systematic uncertainty in each measurement.
The way to estimate and to reduce it is explained in the Section~\ref{sect:m1}.

The width of the peak is in fact the combination of the TOF and 
electronic resolutions.
\begin{equation}
\sigma_{peak}=\sqrt{\sigma_{TOF}^2+\sigma_{electronics}^2}
\label{eq4}
\end{equation}

The electronic resolution $\sigma_{electronics}$ mostly
originates from the TDCs resolution.
It was estimated from the spectrum of $\frac{t_1}{2}$ in which 
the same PM signal generates the common START for the TDC and 
the delayed STOP (Fig.~\ref{fig:scheme}). In fact, 99\% of events 
in this spectrum were located in a single channel.
The resulting value is $\sigma_{electronics} \sim 6.9$ ps.
The corresponding corrections were implemented following Eq.~\ref{eq4}. 

Both methods were verified in simulations. 
The values $t_1$ and $t_2$ were simulated event-by-event 
using random generators. The generator for $t_1$ imitated the 
measured $t_1$ spectrum while the $t_2$ generator was Gaussian.
The width of the Gaussian distribution is equal to $2\sigma_{TOF}$.

The generated $t_1$ and $t_2$ values were then digitized in 
accordance with the discrete scale of LeCroy2228B TDC and recorded
in the same format as the experimental data files. 
The experimental and simulated data were analyzed using the same codes. 
The simulated TOF resolution  was reconstructed with the accuracy of $0.1$ ps
in the case of direct calculation and $0.3$ ps using the Gaussian fit.  
 
One well-known problem of timing measurements is the effect of 
time walks of constant-fraction discriminators (CFD). CFDs generate additional
time shifts (time walks) which depend on pulse heights.
To extract the TOF resolution, the events were always selected
using the criterion  
\begin{equation}
L_{i}\leq \sqrt{(A_1-ped_1)(A_2-ped_2)} \leq L_{i+1}
\end{equation}
\noindent such that the pulse heights $A_1$ and $A_2$ of the selected events
are nearly constant (for example, events from the beam peak). 
Therefore the influence of time walks on  was minimized.

\section{TOF resolution with fine-mesh R7761-70 photomultipliers
in comparison with R2083 PMs}\label{sect:m1}

In this section we report a comparative measurement of the TOF resolutions of 
two similar counters. One counter was equipped with  
fine-mesh R7761-70 photomultipliers, and the other 
with R2083 PMs. Both counters were made of $2\times 3\times 50 cm^3$ 
Bicron408 scintillation bars. The photomultipliers were directly attached to
the bar butts. The diameter of the photocathodes of R7761-70 photomultipliers, 
of 27 mm, covered only $\sim 80\%$ of the butt surfaces of the ends of 
the scintillator bar. Therefore the light collection in this counter
was lower by factor $~0.8$ than in the case of R2083 PMs.  

The beam irradiated the counters at their centers in the
direction perpendicular to the counter axis. 
The TOF resolution of both counters was measured at the same conditions: 
after the beam adjustment the counters were attached one-by-one 
to the collimation system. 
The replacement of the counters took few minutes. During that short break 
the beam was off. However, the beam parameters were kept fixed. 
The beam spectra and the count rates were similar in both measurements
(Fig.~\ref{fig:bsp1}). 
To avoid the effect of the TDC differential non-linearity, 
the TOF resolution of each counter was measured successively six times. 
The scheme of the first measurements is shown in 
Fig.~\ref{fig:scheme}. In the second measurement the signal of that PM which 
does not trigger the acquisition, was additionally delayed for $1$ ns. 
In the third measurement the delay was 2 ns. Then
the PMs were switched and the series of the measurements with additional 
0, 1, and 2 ns delays was repeated.

The data were analyzed in an identical way: 
the same codes was running over the data files corresponding 
to 12 ($6\times 2 $) measurements. Only events from the beam-peak area 
$31.5 MeV \leq L \leq 32.5 MeV$ were selected to extract the TOF resolutions. 

The results were obtained by using both the direct calculation and the fitting
of $(t_1-t_2)/2$ spectra. 
They are summarized in Table~\ref{tab1} and Table~\ref{tab2}. 
The TOF resolution obtained with R7761-70 PMs was 
in addition scaled by factor $\sqrt{0.8}$ (last columns of Table~\ref{tab1},~\ref{tab2}) 
in order to account for 80\% light collection due to the smaller diameter of 
the R7761-70 photocathodes. Such ``effective" TOF resolution corresponds 
to the same number of photoelectrons produced at the R7761-70 and R2083 photocathodes. 
Further the corrected R7761-70 TOF resolution is used for the comparison with R2083 PMs.  

\begin{table}
\begin{tabular}{cccccc}
Measurement&$PM_{trigger}$&Additional&$\sigma_{TOF_{R2083}}$&$\sigma_{TOF_{R7761-70}}$&$\sigma_{TOFcor_{R7761-70}}$ \\
 & &delay (ns)& & & \\
1 & 1 & 0 & 18.85 & 23.84 & 20.74 \\
2 & 1 & 1 & 19.09 & 19.84 & 17.26 \\
3 & 1 & 2 & 16.14 & 19.70 & 17.14 \\
4 & 2 & 0 & 16.30 & 21.68 & 18.86 \\
5 & 2 & 1 & 16.43 & 22.56 & 19.64 \\
6 & 2 & 2 & 17.60 & 18.59 & 16.17 
\end{tabular}
\caption{TOF resolutions (ps) of R2083 and R7761-70 PMs directly calculated 
from the samples of collected events}
\label{tab1} 
\end{table}

\begin{table}
\begin{tabular}{cccccc}
Measurement&$PM_{trigger}$&Additional&$\sigma_{TOF_{R2083}}$&$\sigma_{TOF_{R7761-70}}$&$\sigma_{TOFcor_{R7761-70}}$ \\
 & &delay (ns)& & & \\
1 & 1 & 0 & 18.25 & 22.67 & 20.28 \\
2 & 1 & 1 & 19.02 & 19.67 & 17.50 \\
3 & 1 & 2 & 15.76 & 19.19 & 17.16 \\
4 & 2 & 0 & 16.72 & 20.92 & 18.70 \\
5 & 2 & 1 & 15.97 & 22.00 & 19.68 \\
6 & 2 & 2 & 16.88 & 18.33 & 16.40 
\end{tabular}
\caption{TOF resolutions (ps) of R2083 and R7761-70 PMs obtained using Gaussian fits.}
\label{tab2} 
\end{table}

The statistical error of each measurement was $\sim 0.1$ ps. The systematic uncertainty 
mostly arises from the TDC differential non-linearity. It
can be estimated as the deviation of six measured data points from their mean 
value
   
\begin{eqnarray}
\sigma_{TOF_{mean}}=\frac{1}{6}\sum_{i=1}^{6}\sigma_{TOF_i} \\
\Delta \sigma_{TOF syst}=\sqrt{\frac{1}{6}\sum_{i=1}^{6}(\sigma_{TOF_i}-\sigma_{TOF_{mean}})^2}
\end{eqnarray}

\noindent The estimates for the systematic error in one measurement 
is $\Delta\sigma_{TOF R2083 syst}= 1.21$ ps
and $\Delta\sigma_{TOF R7761-70 syst}= 1.58$ ps for the direct calculation method,
and $\Delta\sigma_{TOF R2083 syst}= 1.17$ ps
and $\Delta\sigma_{TOF R7761-70 syst}= 1.38$ for the Gaussian fit.
The error for the average of 6 measurements should be 
scaled by $\frac{1}{\sqrt{6}}$. The averages for the R2083 resolution and for 
the corrected R7761-70 resolution are
\begin{equation}
\sigma_{TOF R2083}=17.4\pm0.49{\hskip 1cm}
\sigma_{TOF R7761-70}=18.3\pm0.64
\end{equation}
\noindent for the direct calculations and
\begin{equation}
\sigma_{TOF R2083}=17.1\pm0.48{\hskip 1cm}
\sigma_{TOF R7761-70}=18.3\pm0.56
\end{equation}

The ratio of the R7761-70 and R2083 resolutions is
\begin{equation}
\frac{\sigma_{TOF R7761-70}}{\sigma_{TOF R2083}}=1.05\pm0.066
\end{equation}
\noindent and
\begin{equation}
\frac{\sigma_{TOF R7761-70}}{\sigma_{TOF R2083}}=1.07\pm0.062
\end{equation}
for the direct calculation and for the Gaussian fits respectively.
Further the average of both methods
\begin{equation}
\frac{\sigma_{TOF R7761-70}}{\sigma_{TOF R2083}}=1.06\pm0.064
\end{equation}
\noindent is used. This ratio means that, if the number of photoelectrons is the same, 
the TOF resolution with R7761-70 photomultipliers would be $\sim 6.6\pm 6\%$ worse 
than in the case of R2083 PMs.

The result proves the good timing performance of R7761-70 PMs.
Moreover, this ratio was obtained at low 
$HV \sim 1300 V $ of R7761-70 PMs. It is well known that
PM timing properties becomes better at higher HV (see, for example, Ref.~\cite{ru1}).
In addition, as it will be discussed in the section~\ref{sect:mf}, 
$\sigma_{TOF R7761-70}$ might be better in magnetic field.

\section{Dependence of TOF resolution on light output}\label{sect:rlout}

To retrieve the dependence of the TOF resolution on the light output, 
the sample of collected events was divided into 
many sub-samples following the criterion

\begin{equation}
L_{i}\leq \sqrt{(A_1-ped_1)(A_2-ped_2)} \leq L_{i+1}
\end{equation}

The width of the bins $L_{i+1}-L_{i}$ was $1$ MeV.
For each bin the width of the peak $\sigma_{peak}$ was derived using both methods.

\begin{figure}
\begin{center}
\includegraphics[width=14.0cm]{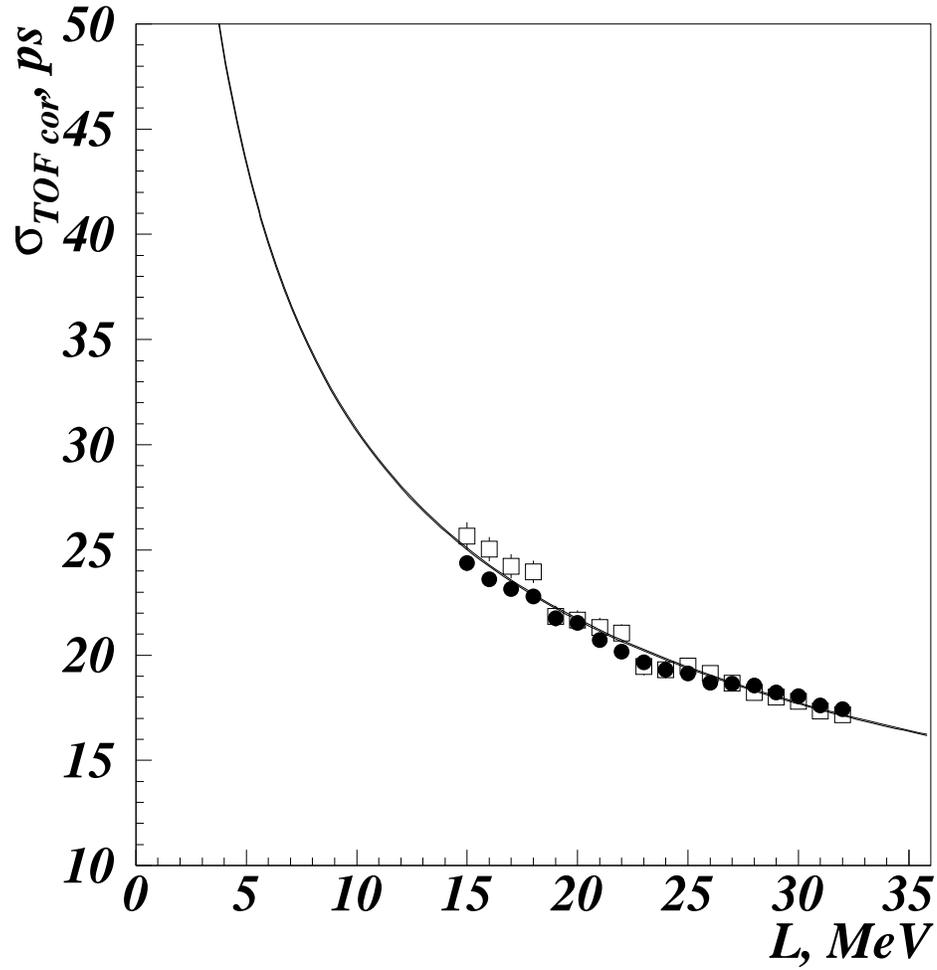}
\caption{Dependence of the measured time-of-flight resolution on the light output.
Circles are directly calculated from the sample of collected events.
Open squares are the results obtained using the Gaussian fit. 
Error bars correspond to statistical errors only.
Where not visible, the errors bars are smaller than the symbol size.
The curve is the fit of experimental data. The difference between two fits
corresponding to each data set is not seen in the plot.}
\label{fig:tres2}
\end{center}
\end{figure}

The results collected with  R7761-70 PMs in the October 2008 
are shown in Fig.~\ref{fig:tres2}.
The data obtained by two methods are consistent. 
The minor deviation at the lower light output is explained by 
the increasing contamination of scattered protons. This background
differently affects the results obtained by each method. 

The data points are well fitted by $\frac{C}{\sqrt{L}}$. The results of the fit are  
\begin{equation}
\sigma_{TOF}(L)=\frac{96.9\pm 0.27_{stat}\pm2_{syst} (ps)}{\sqrt{L(MeV)}} 
\end{equation}
\noindent with $\chi^2=2.4$ for the directly calculated data points and
\begin{equation}
\sigma_{TOF}(L)=\frac{97.2\pm 0.36_{stat}\pm2_{syst} (ps)}{\sqrt{L(MeV)}} 
\end{equation}
\noindent for $\chi^2=0.95$ for the Gaussian fit. The systematic uncertainty of $2$ ps originates from 
the TDC differential non-linearity and from the accuracy in the determination
of the MIP position.

\section{Dependence of TOF resolution on coordinate and track angle}

The TOF resolution is related to the PM timing resolutions
$\sigma_{PM1}$ and $\sigma_{PM2}$ as 
\begin{equation}
\sigma_{TOF}=\frac{1}{\sqrt{2}}\sqrt{\sigma_{PM1}^2+\sigma_{PM2}^2}
\end{equation}
\noindent where $\sigma_{PM1}$ and $\sigma_{PM2}$ are the effective
resolutions in each PM channel. If a scintillation occurs in the middle of a counter, the numbers
of light photons that reaches each PM, are equal. 
Therefore $\sigma_{PM1}=\sigma_{PM2}$.
If a scintillation is located near a counter end $\sigma_{PM1}\neq \sigma_{PM2}$.
Due to the exponential light attenuation inside a counter
the number of photons which reach each photomultiplier is different. 
This effect may generate the dependence of the TOF resolution on 
the axis along the counter axis.

\begin{figure}
\begin{center}
\includegraphics[width=7.0cm]{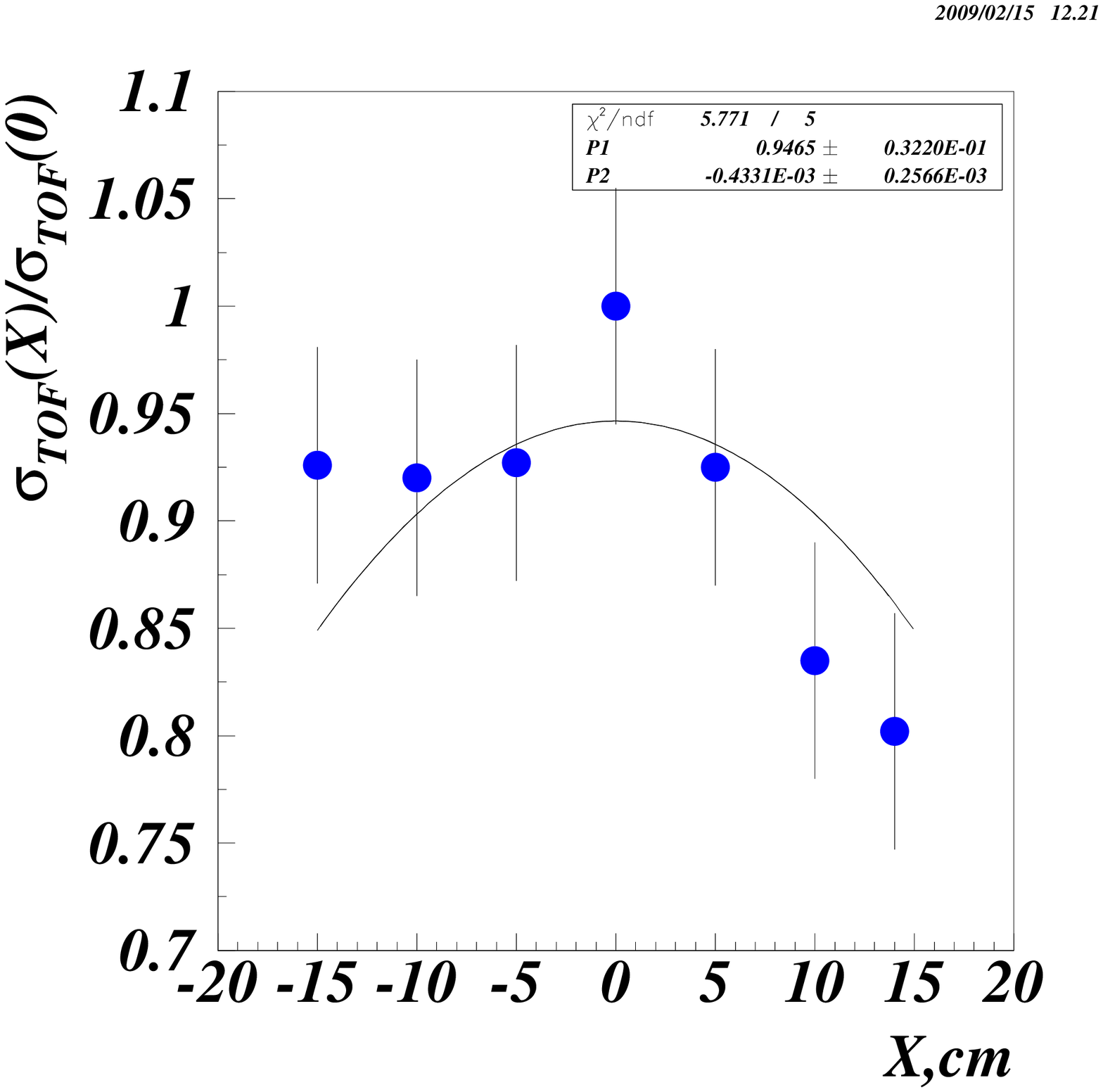}
\caption{Dependence of the TOF resolution on the coordinate along the counter axis.
}
\label{fig:xcoor}
\end{center}
\end{figure}

To retrieve this dependence, the TOF resolution
was measured at different coordinates along 
the counter axis.
The results are shown in Fig.~\ref{fig:xcoor}. The TOF resolution
is $\sim 15\%$ better near the counter ends.
 
In addition, two measurements were performed with the beam directed 
at $60^{\circ}$ relative the counter axis. Within the systematic uncertainty,
the TOF resolution was the same as that obtained with
the perpendicular beam. More accurate
data for the $\sigma_{TOF}$ dependence on the coordinate and the angle
will be obtained during next beam runs.

\section{Operation of R7761-70 PMs in magnetic field}\label{sect:mf}

In this section we present first results on the operation
of R7761-70 photomultipliers in magnetic field. 
Both photomultipliers were placed inside 
air-cooled solenoids. The solenoids were designed and manufactured by the 
KNU group. They comprise two parallel sections each wounded with 
8 layers of 1-mm dia cooper wire. The 5 mm gap between the sections allows 
the air passage in order to improve the cooling efficiency. In future the number 
of sections will be increased to 4.

Each section generates 100 Gauss of the magnetic field per 1 A of the current.
Due to heating, the maximum current is limited to $\sim 6$ A. Accordingly, the upper 
limit of the magnetic field in the two-sectional solenoid is 1200 Gauss. 
Significant increase of the generated field should be expected 
if the air cooling would be replaced with the water-cooling system. 

\begin{figure}
\begin{center}
\includegraphics[width=10 cm]{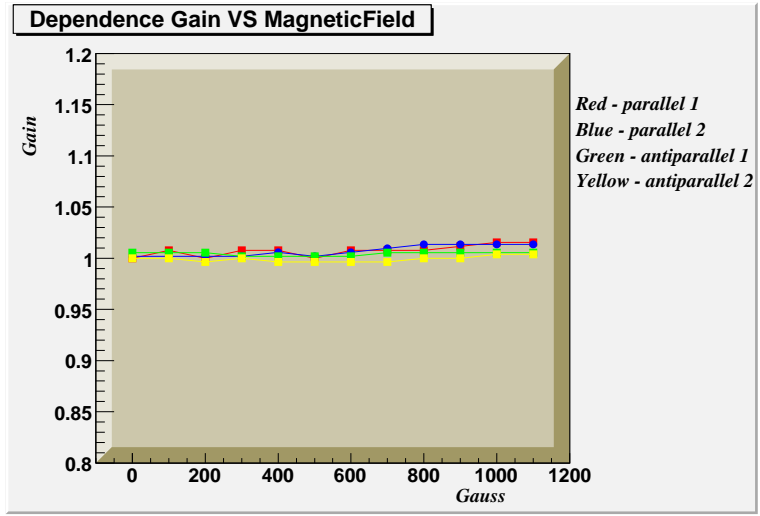}
\caption{Dependence of R7761-70 gain on magnetic field. }
\label{fig:mf4}
\end{center}
\end{figure}

\begin{figure}
\begin{center}
\includegraphics[width=10 cm]{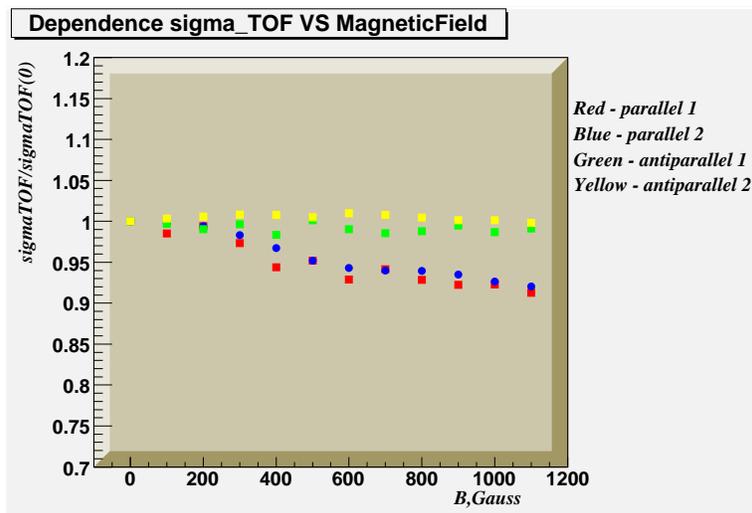}
\caption{Dependence of R7761-70 TOF resolution on magnetic field.}
\label{fig:mf5}
\end{center}
\end{figure}

The measurements were carried out first at 0 Gauss, then at 1100, 1000, 900 ...100, 
and again at 0 Gauss.  All the data takings were done at the same conditions 
one-by-one. The field was varied by changing remotely the solenoid currents.

In total, 4 series of data were taken. In the first series
the direction of the magnetic was chosen parallel to the PM axis.
In the second series the signal of one PM was additionally delayed for 1 ns.
Then the direction of the magnetic field was turned to 180 deg
by switching the polarity of the power supply. After that the 3rd and 4th 
series were done in the similar way.

The results are shown in Fig.~\ref{fig:mf4} and Fig.~\ref{fig:mf5}. The gain of 
R7761-70 PMs was attributed to the position of the beam peak. The beam spectrum 
in this measurement is shown in the panel 6 of Fig.~\ref{fig:badj}.
No any dependence of the gain (i.e. shift in the peak position)
on the magnetic field was observed.

Surprisingly, the TOF resolution becomes better if 
the magnetic field is oriented parallel
to the PM axis. The effect reaches $\sim8\%$ at 1100 Gauss.
The antiparallel field does not affect the TOF resolution. This observation 
has to be checked in next beam runs.

\section{CTOF Estimates} 

In real CTOF assembly most of fast particles will be detected at 
forward angles. Below we present  the estimates
of the expected TOF resolution for minimum-ionizing particles
emitted from a target at $90^{\circ}$ and $45^{\circ}$.

A minimum-ionizing particles passing through
a 3-cm thick counter perpendicular to its axis, deposit $\sim 6$ MeV of energy 
(Fig.~\ref{fig:calibr}).  
The corresponding TOF resolutions are $\sigma_{TOF_{R2083}}=17.3\sqrt{31.5/6}\approx 39.6$ ps
and $\sigma_{TOF_{R7761-70}}=18.3\sqrt{31.5/6}\approx 41.7$ ps
for R2083 and R7761-70 PMs respectively.

The scintillation bars in the CTOF assembly will be viewed through 
light guides. The long bent light guides 
in the CTOF design with ordinary R2083 PMs (Fig.~\ref{fig:des1}) 
deliver $\sim 30\%$ of scintillation light~\cite{tdr}. The 
estimate for the expected CTOF resolution with R2083 PMs would be
\begin{equation}
\sigma_{TOF R2083 LG}(90^{\circ})=\sqrt{\frac{1}{0.3}}\times39.6 ps \approx 72.3 ps 
\end{equation}

In the CTOF design with fine-mesh photomultipliers (Fig.~\ref{fig:des1}), the light guides
will shorter and not bent. One may assume their light transfer efficiency 
around $50\%$. The corresponding estimate  
at $90^{\circ}$ is
\begin{equation}
\sigma_{TOF R7761-70 LG}(90^{\circ})=\sqrt{\frac{1}{0.5}}\times41.7 ps \approx 59.0 ps 
\end{equation}
\noindent It is essentially better than that number for R2083 PMs.

Minimum-ionizing particles emitted at $45^{\circ}$, 
deposit $\sqrt{2}$ times more energy in the CTOF counters than 
ones emitted at $90^{\circ}$.
The expected CTOF resolutions are
\begin{eqnarray}
\sigma_{TOF R2083 LG}(45^{\circ})=\\
\nonumber{0.85\frac{1}{\sqrt{2}\sqrt{0.3}}\times39.6 ps \approx 43.5 ps} \\
\sigma_{TOF R7761-70 LG}(45^{\circ})=\\
\nonumber{0.85\frac{1}{\sqrt{2}\sqrt{0.5}}\times41.7 ps \approx 35.4 ps}
\end{eqnarray}
\noindent where the factor $0.85$ originates from the dependence 
of the TOF resolution on the coordinate. 
\begin{figure}
\includegraphics[width=6.7cm]{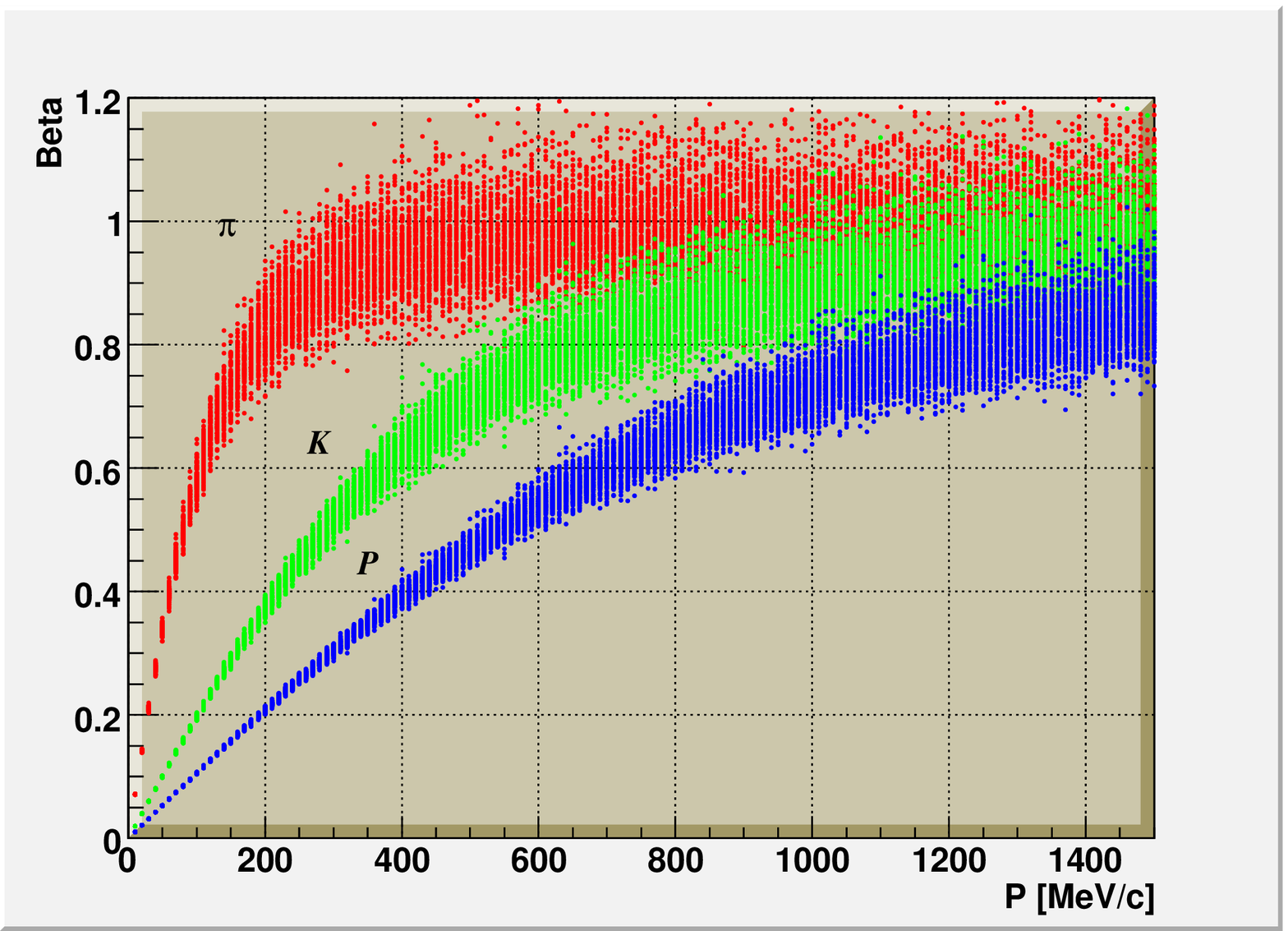}
\includegraphics[width=6.7cm]{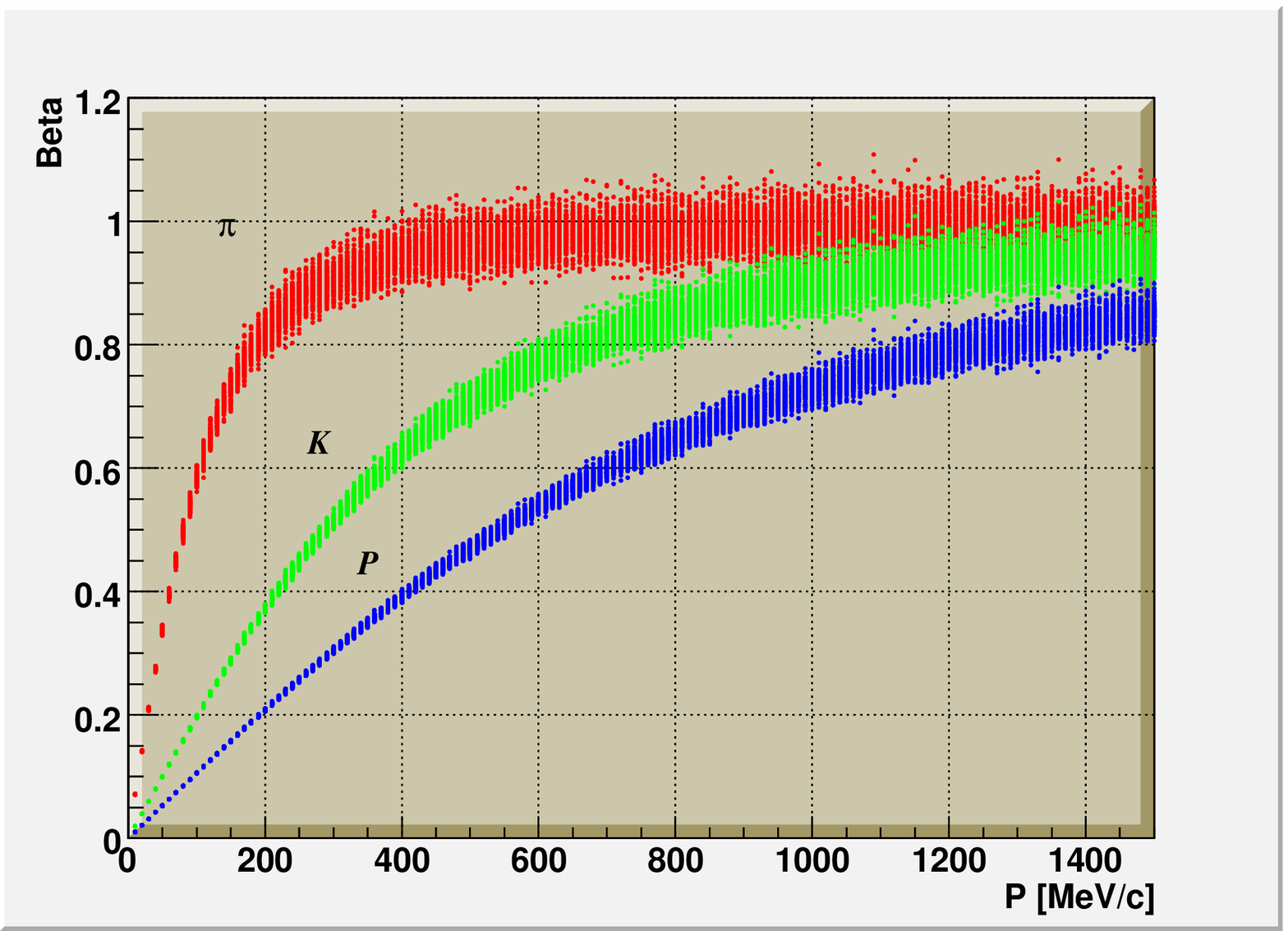}
\caption{Simulated particle identification in the CTOF detector with fine-mesh
R7761-70 photomultipliers at $90^{\circ}$(left) and $45^{\circ}$(right). }
\label{fig:id1}
\end{figure}

The simulated particle identification (PID) based on the quoted above estimates 
is shown in Fig.~\ref{fig:id1}. The particle speed $\beta$ is used as an identification
parameter  
\begin{equation}
\beta=\frac{l_{TOF}}{c\times TOF},
\end{equation} 
\noindent where $l_{TOF}$ is a distance from a vertex point inside a target to 
a hit point on a counter surface. 
The simulations include four sources of uncertainties:\\
i) the TOF resolution and its dependence of the light output; \\
ii) the coordinate resolution $\sigma_x=\sigma_{TOF}\times15cm/ns$;\\ 
iii) CTOF azimuthal granularity;\\
iv) The reconstruction of the vertex point $\pm 1$ mm. 

At $90^{\circ}$ kaons are discriminated from pions up to 
$\sim 600$ MeV/c. Protons are discriminated from kaons up to $\sim 1000$ MeV/c.
In the most critical region near $45^{\circ}$ PID is essentially better:
protons are separated from kaons up to $\sim1400$ MeV/c, and
pions from kaons up to $\sim850$ MeV/c.

\section{Summary and future R\&Ds at KNU.}  

The measured ratio of the effective TOF/timing resolutions of R7761-70 and
R2083 PMs $\frac{\sigma_{TOF R7761-70}}{\sigma_{TOF R2083}}=1.06\pm0.064$ 
proves the advantages of the CTOF design with fine-mesh photomultipliers.
This design will be more simple, less expensive, and will provide better performance
than the ``conservative" design with ordinary R2083 PMs. 

More information on the properties of fine-mesh photomultipliers will 
be obtained in the next beam runs. The R\&D program  includes:\\
- measurements of the relative TOF resolution of several counters
equipped with R7761-70, R5924-70 and R2083 PMs;\\
- operation of  R7761-70 and R5924-70 photomultipliers 
in magnetic field up to 0.4 Tesla;\\
- accurate measurements of the dependence of the TOF resolution 
on the coordinate and track angle.

At the next stage silicon photomultipliers, 
micro-channel plates could be tested as well using the same technique. 
Another option might be the development of a time-of-flight system 
for the detection of neutrons. The new neutron beam line is currently 
under construction at the MC50 Cyclotron.
It will offer a tool for testing prototype
counters of the CLAS12 neutron detector~\cite{silv}.

It is a pleasure to thank Dave Kashy, 
Cyrill Wiggins, and Volker Burkert for 
the fruitful and efficient collaboration. Silvia Nicolai helped 
to present the KNU results at JLAB. 
Our special thanks to the staff of the MC50 Cyclotron for their 
assistance during beam runs, and especially 
June Yong Han for the assistance in solving the problem of 
the beam tuning. This work was supported by Korea Science and Engineering 
Foundation.

\end{document}